# Digital Calibration Method for High Resolution in Analog/RF Designs

Submitted in partial fulfillment of the requirements for

the degree of

Doctor of Philosophy

in

Electrical and Computer Engineering

Renzhi Liu

B.S., Electronics Engineering and Computer Science, Peking University

Carnegie Mellon University

Pittsburgh, PA

July, 2015

*To my dear wife, Liang*

# Acknowledgments

First, I would like to thank my advisor Professor Larry Pileggi for all his support and continuous guidance throughout the past five years. He opened up a wide variety of research topics that I can explore during my early years at Carnegie Mellon University (CMU) and helped me find my research path that eventually leads to this dissertation.

Second, I would like to thank my co-advisor Professor Jeff Weldon. It was him who guided me to the RF circuit design in this dissertation, during which I gradually came up with the proposed digital calibration method.

Next, I would like to thank the rest of my dissertation committee members. Professor Jeyanandh Paramesh gave the lectures of the most valuable class to me at CMU, which was radio frequency integrated circuits design. This class shaped my research interests as well as my future career path. Dr. Gokce Keskin proposed the original design method on which my new method was based. His work inspired me during my graduate study and I became an early adopter of his proposed design method in one of my early chip design.

I would also like to thank my fellow lab members. Kaushik Vaidyanathan has patiently taught me, an analog circuit designer, how to do proper semi-custom digital circuit design. As a senior student to me, his advices also helped me to go through graduate student life as well as to find my career path. David Bromberg, Ekin Sumbul, Vanessa Chen and Vehbi Calyir entered the Ph.D. program at the same year as I did. I appreciate all the collaborations we have in classes as well as in research projects with the thoughtful discussions and sincere advices. I would also like to thank the rest of the current and past lab members in Professor Pileggi's group who overlapped with me: Andrew Phelps, Bishnu P Das, Cheng-Yuan Wen, Curtis Ratzlaff, Dan

Morris, Fazle Sadi, Jinglin Xu, Kuntal Shah, Meric Isgenc, Ozan Iskilibli, Qiuling Zhu, Rongye Shi, Shaolong Liu, Soner Yaldiz and Thomas Jackson. I treasure the years that we spent together at CMU.

The work presented in this dissertation was supported in part by the Intelligence Advanced Research Program Agency and Space and Naval War-fare Systems Center Pacific under Contract No. N66001-12-C-2008.

Last but not least, I would like to thank my parents. They have raised me as an individual that desires for knowledge and truth. I will be forever grateful for all the things they taught me and the characters they gave me.

Finally, I would like to give my deepest thanks to my beloved wife Liang Tang, although words cannot describe the magnitude of my gratitude here. It was a small probability event that we met each other at Peking University and then came to CMU together as graduate students. I cannot achieve anything without your help, your love and your sacrifice.

# Abstract

Transistor random mismatch continuously poses challenges for analog/RF circuit design for achieving high accuracy and high yield as the process technology advances. Existing statistical element selection (SES) design method can improve transistor matching property, but it falls short of being a general calibration method due to its limited calibration range.

In this dissertation, we propose a high resolution digital calibration method, called extended statistical element selection (ESES). As compared to the SES method, the ESES method not only provides wider calibration range, but also it results in higher calibration yield with same calibration resolution target. Two types of ESES based calibration application in analog/RF circuits are also proposed. One is current source calibration and the other is phase/delay calibration.

To verify this proposed digital calibration method in circuit implementation, we designed, fabricated and tested a wideband harmonic rejection receiver design. The receiver utilizes ESES-based gain and phase error calibration for improving harmonic rejection ratios. With the high calibration resolution provided by the ESES method, after calibration, we achieved best-in-class harmonic rejection ratios. To extend the application of the proposed method, we further designed a current-steering D/A data converter (CS-DAC). The CS-DAC utilizes ESES-based amplitude and timing error calibration for improving linearity performance. Simulation results showed that we can achieve more than one order of magnitude linearity improvement after performing ESES-based calibration in the CS-DAC.

# Contents

# List of Figures

# List of Tables

# 1 Introduction

As CMOS technology continues to scale [1], digital circuits have leveraged the corresponding technology improvements to achieve better area, speed and power. In contrast, analog circuits have scaled sub-optimally with process technology. One important limiter is transistor random mismatch in CMOS technology that degrades the analog circuit performance, specifically accuracy and yield [2].

To combat the impact of transistor random mismatch on analog circuit performance, a straightforward method is to increase transistor sizes to improve matching properties. However, this is not favored since this method increases analog circuit area and also lowers circuit bandwidth.

A novel form of sizing based on redundancy at the subcomponent level, statistical element selection (SES), was proposed in [3] to address this challenge. The SES method provides combinatorial design choices to be digitally selected from among a large population, thus achieving excellent matching properties after selection. It has been shown that this combinatorial redundancy can be applied to analog designs for high resolution calibration, such as for tuning input offset of a differential pair [3]. However, the calibration tuning range of the SES method is purely determined by transistor random mismatch and this limitation of the tuning range makes SES methodology unsuitable for many calibration applications as a general calibration method.

To address this limitation and make the calibration range a design parameter, an extended statistical element selection (ESES) method is proposed here that is based on non-uniform sizing of the elements under selection. We have applied the ESES-based high resolution calibration method to two categories of calibration applications for analog and RF circuits, current calibration and phase delay calibration. Our comparison of traditional calibrations methods for these applications indicates that our high resolution digital calibration method has lower calibration circuit overhead, improved calibration resolution and flexible circuit implementation.

As a demonstration for the proposed high resolution digital calibration method, we have designed a wideband harmonic rejection receiver using the ESES methodology. The ESES-based calibration is applied to correct both gain errors and phase errors. By having high calibration resolution, best-in-class $2^{nd}$ to $6^{th}$ order harmonic rejection ratios are achieved. Moreover, by calibrating gain errors and phase errors independently, both $3^{rd}$ order and $5^{th}$ order harmonic rejection ratios can be optimized simultaneously.

As another example, for the proposed high resolution digital calibration method we designed a current-steering D/A data converter (CS-DAC). We applied ESES-based current and phase delay calibration method to the CS-DAC design to correct amplitude error as well as timing errors of the thermometer coded MSBs' current cells in CS-DAC. After calibration, we can achieve much improved linearity performance.

This document is arranged as follows:

- Chapter 2 discusses the challenges for analog circuit design with the presence of transistor random mismatch. It surveys existing design methods to address the matching problems, including transistor sizing, SES method and traditional calibration methods.

- Chapter 3 describes our proposed high resolution digital calibration method, known as ESES, and compares it to the original SES method. Two types of calibration application based on ESES method are also described in this chapter.

- Chapter 4 presents a demonstration for the proposed ESES method by a design of harmonic rejection receiver. ESES based calibration method is applied to the receiver design to achieve best-in-class harmonic rejection ratio. The implementation and testing of the design is detailed in this chapter.

- Chapter 5 presents a second example for the proposed ESES method by a design of CS-DAC. The amplitude and timing errors of the data converter are calibrated following ESES-based methods to achieve greatly improved linearity performance of the CS-DAC.

- Chapter 6 concludes with a discussion of key results as well as future research direction.

# 2   Background

## 2.1   Random Mismatch in CMOS Technology

Transistor random mismatch poses a significant challenge for analog circuit design. As pointed out in [2], the matching properties of CMOS transistors, especially the threshold voltage mismatch, determine the yield and performance of many data converters. For the application of parallel analog signal processing, random mismatch can determine the equality of parallel signal paths, which has a great impact on the system performance.

With CMOS scaling, the matching properties of transistors become worse. As shown in [4], from technology node of 180 nm to 45 nm, the transistor feature size has scaled down by 4X, while the threshold voltage matching coefficient has only improved roughly by 2X. According to Pelgrom's transistor mismatch model [5], this would lead to an increase in threshold voltage mismatch if analog transistor sizes scale proportionally with process technology.

For sub-20nm CMOS, the FinFET has been the device of choice for volume production. The threshold voltage matching coefficient of the FinFET is shown to be low in [4] since the channel doping (a major source of variability in threshold voltage random mismatch) can be absent in fully depleted devices. However, line-edge roughness of the fins degrades the threshold voltage matching dramatically for deeply scaled nanoscale processes [6]. Moreover, the assumption of absence of channel doping may not hold true if channel doping is ultimately used for generating multiple threshold voltage flavors instead of having different gate work-functions, since the latter has significantly higher process complexity and cost [7]. Hence, analog circuit design will continue to suffer from transistor matching properties with FinFETs.

## 2.2 Transistor sizing for Matching

Pelgrom's transistor mismatch model [5] shows that the standard deviations of the current factor and threshold voltage random variation are proportional to $1/\sqrt{WL}$. So in order to improve matching, one straightforward method is to increase the transistor sizes W and/or L. However, this trade-off between area and matching has very low efficiency as the matching property improvement factor is only a square root of the area increasing factor. And increased transistor gate area is not favored as it results in increased gate capacitance and parasitic capacitance that decrease circuit speed and performance.

One example of relying on the intrinsic accuracy through Pelgrom type transistor sizing is for a high resolution current steering DAC (CS-DAC) design. In [8], a 14-bit CS-DAC was designed with each current cell sized sufficiently big enough to ensure certain matching property for achieving 14-bit static linearity. Area of the current cell array totaled more than 5 mm$^2$ in 0.5 μm CMOS technology, while dynamic performance of the CS-DAC greatly suffered from the resulting parasitic capacitance. Since the transistor matching property hasn't improved at the same pace as the process technology scaling, the CS-DAC design using Pelgrom type transistor sizing would produce proportionally larger analog circuit area, which mainly consists of overly sized current cells for matching purpose as technology process scales.

## 2.3  Statistical Element Selection (SES) Design Methodology

Matching properties can be improved by redundancy. A traditional method to create redundancy is to have a large population of cells to choose from. Instead of only having one cell, redundancy is based on having N cells and choosing the best one from that population. Assuming these N cells follow the same independent distribution and the success rate of each cell passing certain specifications is $P_{success}$, then the total success rate of having N cells available for selection is $1-(1-P_{success})^N$, which is significant improvement over $P_{success}$. This method was incorporated in the flash ADC design in [9] to improve ADC linearity. Although traditional redundancy can improve the yield of the analog circuit design, the cost is very high (scales by N).

With statistical element selection (SES) that was proposed in [3], the approach is based on digitally selecting K elements from a set of N elements and sacrificing N-K elements, resulting in a combinatorial redundancy. The number of available combinations increases exponentially as N and K increase, enabling an exponential number of design choices that can be digitally selected. By having this combinatorial redundancy, the failure rate to pass certain specifications can be improved by orders of magnitude. Keskin et. al. demonstrated multiple orders of magnitude lowering of standard deviation of the comparator input offset voltage after optimal selection [10]. Therefore, the total area requirement for meeting certain matching properties is lowered dramatically, thereby consuming much smaller area, even when considering the sacrificial area that enables combinatorial redundancy. This design methodology can solve the transistor matching problem at a relatively low circuit overhead (usually a digital controller), as demonstrated in a 14-bit current-steering D/A data converter design in [11]. And since there exists an exponential number of design choices aggregated around the nominal design value, this methodology also enables high resolution calibration within a small range.

The SES design method has also been regarded as an efficient design method for FinFET technologies [12] [13] [14] as well. Furthermore, the design method has been employed for trusted IC manufacturing with split-fabrication for enhanced IC security [15] [16].

## 2.4   Calibration Methods for Current Sources and Phase Delays

### 2.4.1   Current Sources Calibration Methods

To address the transistor random mismatch problem for current sources, many calibration methods have been proposed. Current adjustment can be done by varying the bias voltage of the current source [17] [18]. By adding a serial resistor to the gate of the current source and supplying a small tunable current flowing through that resistor, the effective biasing gate voltage of the current source can be tunable. The current calibration circuit is shown in Figure 2.1. The major overhead of this method is the calibration DAC (CALDAC) and associated current mirror structure that can require significant area, as shown in [18].

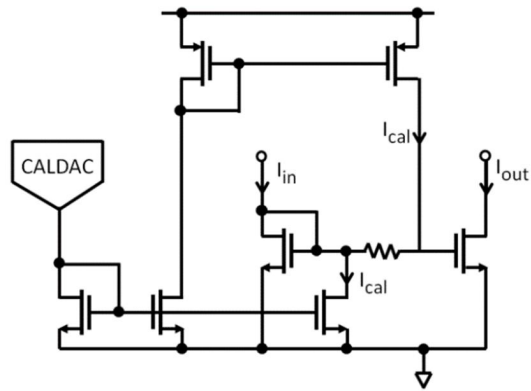

Figure 2.1 Current calibration circuit based on varying gate voltage [18].

More directly, current adjustment can be done by adding small tunable current source in parallel to the current source under calibration. There are mainly two methods to create the small tunable current source. The first method is to tune the gate voltage of the added current source and use a capacitor to store the calibrated gate voltage value for use during normal operation [19] [20]. A conceptual diagram of this method is shown in Figure 2.2 (a). This method can share the CALDAC among all current sources under calibration. However, it also requires frequent refreshing of the capacitor to handle charge leakage. Moreover, the resulting calibration schemes for this method presented in [19] [20] involve high analog circuit overhead that also suffers with process scaling. Another method is to attach the CALDAC current output directly in parallel to the current source under calibration [21] [22]. A conceptual circuit diagram of this method is shown in Figure 2.2 (b), which is more digital calibration orientated, but each current source needs one CALDAC. For high calibration accuracy, a very small least significant bit (LSB) current cell of CALDAC is needed for a small calibration step, which usually leads to LSB current cell design with much larger channel length than that of the current source under calibration. As a result, the area overhead for the CALDACs is observed to be high in [21] [22].

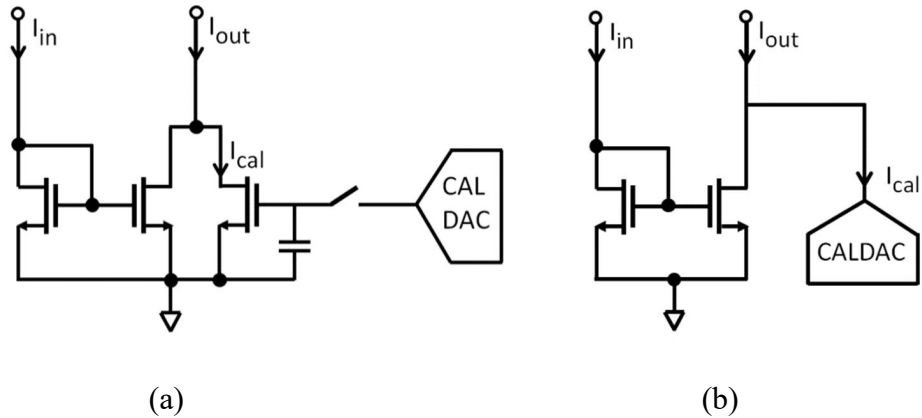

(a) (b)

Figure 2.2 Conceptual circuit diagram of current calibration based on attaching extra current source (a) method one (b) method two.

### 2.4.2 Phase Delay Calibration Methods

Phase delay matching also suffers from transistor random mismatch. Most of the phase delay calibration methods utilize one of the two existing tuning mechanisms for delay element: current-starved inverter [23] and shunt-capacitor inverter [24]. These techniques were proposed originally for digital delay-locked loops or digital phase-locked loops design with large tuning range. But they can also be modified to have finer tuning resolution to be applied to analog/RF designs where digital signals when high timing requirements are needed; e.g. timing skew requirement of time-interleaved ADC and phase offset requirement of multi-phase local oscillator (LO).

The current-starved inverter with analog control voltage [23] is shown in Figure 2.3 (a). The maximum current available to the inverter is controlled by the control voltage $V_{ctrl}$. By tuning this voltage, the delay become tunable and thus can be calibrated. A digital adaptation [25] is shown in Figure 2.3 (b), with current-starving transistors discretized and hence controlled digitally.

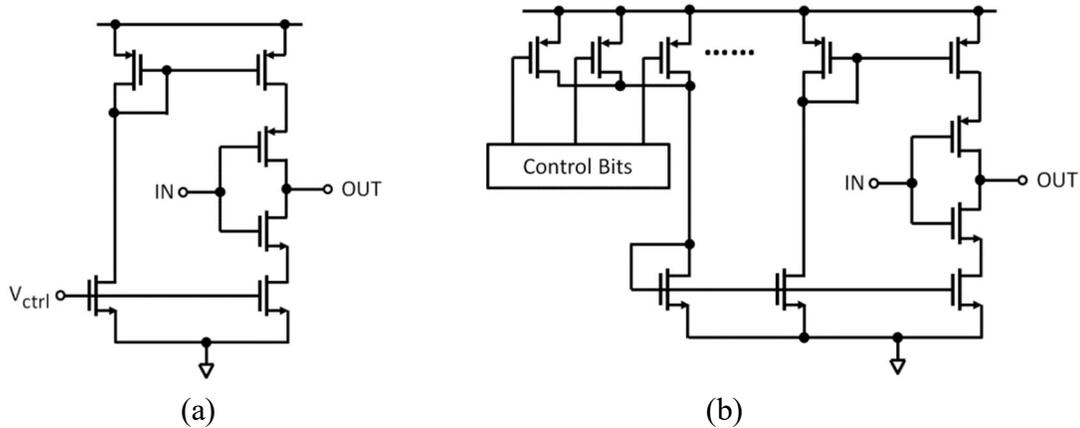

(a)                        (b)

Figure 2.3 Current-starved inverter: (a) analog implementation; (b) digital implementation.

A shunt-capacitor inverter with analog control voltage [24] is shown in Figure 2.4 (a). An MOS capacitor is loading the inverter in series with one transistor. This transistor has tunable "on" resistance by varying control voltage $V_{ctrl}$, thus affecting the effective loading of the inverter and making the delay tunable. A digital adaptation shown in Figure 2.4 (b) was proposed in [26]. The idea is to break the big MOS capacitor into small capacitors in parallel, and tune the loading capacitance by digital control of the MOS capacitor's gate voltage.

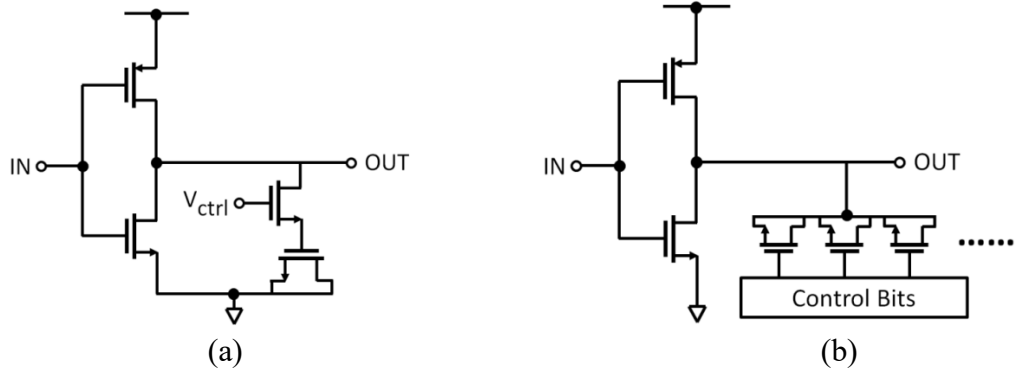

Figure 2.4 Shunt-capacitor inverter: (a) analog implementation; (b) digital implementation.

The analog versions of these two calibration mechanisms involve high analog circuit overhead for generating the control voltage. Although the digital adaptations have much less circuit overhead, the calibration resolution is limited by the smallest current starving transistor or the smallest loading capacitor.

Another limitation of these calibration methods is that the output rising edge and falling edge cannot be tuned separately. This further limits their utility for applications where both edges are critical and need to be calibrated independently.

## 2.5 Summary

In this chapter, we reviewed the matching challenges for the analog/RF circuit design in advanced CMOS technology node and a traditional upsizing technique that can improve transistor matching properties. A more efficient way to improve matching is through combinatorial redundancy and the associated design methodology is called statistical element selection. This method can dramatically improve the yield of the analog circuit design. Traditional calibration methods for current sources and phase delay were also reviewed in this chapter.

# 3  High Resolution Calibration Approach

The original SES methodology enables high resolution calibration within a small range determined by its own random variations. However, the very limited and uncontrollable calibration range makes it unsuitable for many applications. We propose an extended statistical element selection approach that provides for a wider calibration range that can be controlled during the design process. With these features, this new approach can be used as a general calibration method for high resolution purpose. Two types of calibration based on the new design methods are further proposed, which are current calibration and phase delay calibration.

## 3.1  Extended Statistical Element Selection (ESES)

### 3.1.1  Limitations of SES

For the original SES methodology the distribution of the combined K-element is created by the random variation of the equally-sized N elements [3], and a large number of combinations aggregate around the center of the design choices distribution. The clustering effect of SES creates an area of ultra-high density of combinations as available design choices for selection, and therefore, high calibration resolution can be realized around the nominal design value.

However, once the calibration target window deviates from the nominal design value, the calibration success rate drops rapidly. This is because distribution density of the available design choices decreases dramatically as the target window deviates from the center; therefore much less design choices are available for selection in that area, resulting in lowered probability of finding a design choice falling in the target window.

Moreover, the entire calibration range created by the SES method purely depends on the random variation of the elements. The only way to change the calibration range is to purposely change the sizes of the equally-sized N elements, thereby changing their random variations. But this might not be favored as this would also change the circuit performance. In short, the calibration range of the SES method cannot be set independently.

These limitations of the original SES method do not create a big problem if the equally-sized N elements are the only variation source or the strongly dominant variation source in the design. The calibrated result of the combined K-element has to cancel out the impact from other variations sources in the design. If all other variation sources are negligible, the calibration target of the combined K-element can be bounded in a very small region around the nominal design value. Also there is no need for a larger calibration range in this case to cover any outlying calibration targets. As one example, a comparator design with SES-based input offset voltage calibration was shown in [3]. The input differential pair is the dominant variation source for the input offset voltage. Therefore, applying the SES method on the input differential pair can successfully calibrate the input offset voltage to be minimal.

Once there are other non-negligible variation sources, other comparable variation sources or even other dominant variation sources in the design, the calibration target of the combined K-element is determined by the random variation of the other sources. The location of the calibration target can be far away from the center of the design choices distribution or simply out of the calibration range created by the random variations of the equally-sized N elements. In these cases, SES method would be less effective or even not applicable.

In summary, the original SES methods has limited tuning range to counter other variation sources in the design. Without involving another level of coarse calibration preceding the SES based calibration, it has limited application as a general calibration method.

### 3.1.2 ESES Method Overview

To increase the tuning range and to accommodate other dominant variation sources, we propose a design method called extended statistical element selection (ESES) [27]. The ESES design method has non-uniformly-sized N elements. For example, the N elements can be sized as an arithmetic sequence. Meanwhile, still K elements are selected out of the N elements, providing the same combinatorial redundancy as the original SES method. For this ESES method, the overall size of the combined K-element already spans a range for the nominal case, and this range can be controlled by the nominal sizes of the elements as a design parameter. In the case of using arithmetic sequence for the sizes, the distribution range in the nominal case can be controlled by the common difference of the arithmetic sequence while keeping the center value unchanged. With random variation considered in, the distribution of overall size of combined K-element creates a wider calibration tuning range. This proposed ESES method will effectively trade high distribution density around the center of the design choices distribution for a wider tuning range.

As an example, we apply combinatorial redundancy to transistor width by breaking a large transistor into multiple segments and selecting a subset of the segments. The distributions of all available SES/ESES design choices are shown in Figure 3.2. In this example, both SES and ESES methods have parameters of $N = 12$ and $K = 6$. This generates 924 available design choices with each one having 6 segments selected. The nominal transistor width for each segment is set as 1 μm for SES method; and for ESES method, nominal widths are set as an arithmetic sequence of 12 numbers centered at 1 μm and with a gap of 0.02 μm for two adjacent segments (i.e., an arithmetic sequence of 0.89 μm, 0.91 μm, …, 1.09 μm, 1.11 μm with a common difference of 0.02μm). Although there are slight nominal sizing differences, the standard deviation of all segments are set as 0.01 μm. The histogram bin width in Figure 3.2 is set as 0.004 μm. As shown in Figure 3.2 (a), without random variation considered in, all design choices of SES are the same therefore forming a single bin in the distribution. Figure 3.2 (b) shows that with ESES method, even without random variation, the 924 design choices create a wide distribution range. With random variation considered, Figure 3.2 (a) shows that SES method creates a relatively small calibration range, which is purely created by the segments' own random variation. However in Figure 3.2 (b), the ESES method creates a much wider distribution that roughly follows the shape as in Figure 3.2 (b), of which the shape is pre-determined by nominal transistor sizing.

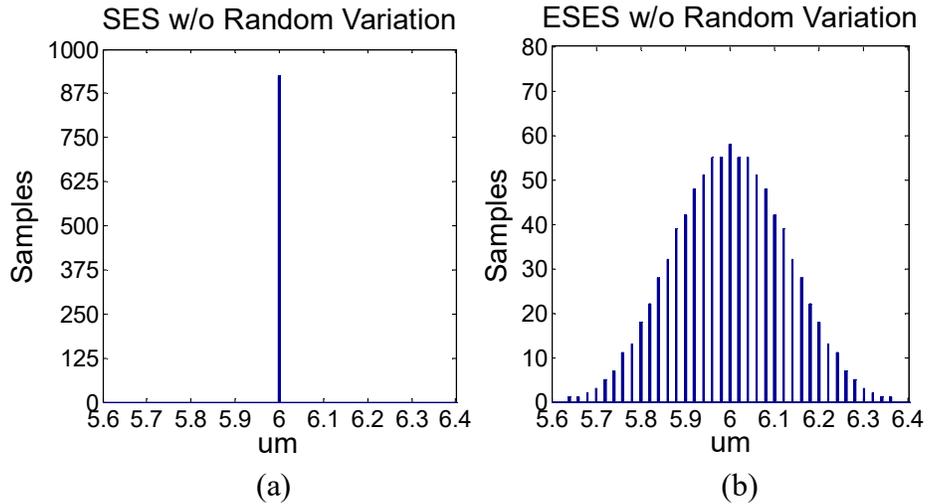

Figure 3.1 Design choices distributions w/o random variation: (a) SES method (b) ESES method.

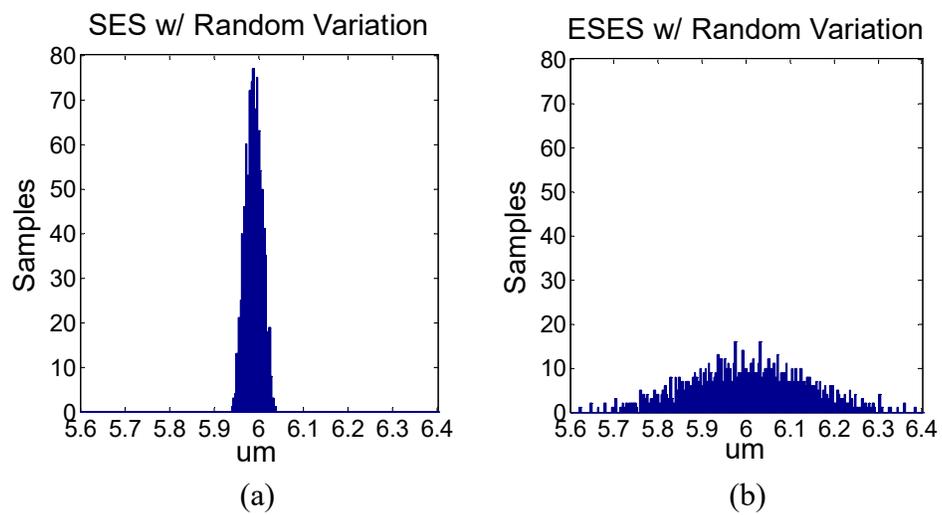

Figure 3.2 Design choices distributions w/ random variation: (a) SES method (b) ESES method.

### 3.1.3 ESES Calibration Resolution and Calibration Range Study

In this section we study the calibration resolution and calibration range of the ESES design method with the comparison to the SES design method. In this study, both SES and ESES methods have parameters of N = 12 and K = 6. As shown in [3], different N, K settings have different calibration yield performance, calibration time (due to different sizes of searching space) and utilization ratios. A setting of N = 12 and K = 6 is picked here as an example to demonstrate the advantages of the ESES method over the SES method, while the following experiments can also be applied to different N, K settings while arriving similar conclusions drawn in this section.

For the SES method, all elements have a nominal size of 1 μm and the standard deviation is set as 0.01 μm. Hence, the size of a combined K-element has a nominal value of 6 μm and a standard deviation of √6 * 0.01 μm = 0.0245 μm. We denote the standard deviation of the subset that has K elements as $\sigma_k$ and this value is used as a unit for normalizing calibration resolution and calibration range. For the ESES method, the nominal sizes of the 12 elements are set as an arithmetic sequence. The average value of the arithmetic sequence is denoted as $a_{ESES}$, which is set as 1 μm if not otherwise stated. The common difference of the arithmetic sequence is denoted as $d_{ESES}$. The standard deviations of each element in the ESES method are calculated based on the assumption that the standard deviation of each element is proportional to the square root of its nominal size. The center value of the standard deviations of the 12 elements is also set as 0.01 μm for fair SES/ESES comparison purpose. For each experiment, a calibration target window size, denoted as $T_{window}$, is set in the unit of $\sigma_k$. This $T_{window}$ value shows the calibration resolution with respect to the standard deviation of the K-element. Once the overall size of the subset (the K elements that are selected) falls into that target window, the calibration process is marked as successful. If we assume the combination falling into the target window follows a uniform distribution, then after calibration the standard deviation of the selected subset decreases to $T_{window}/\sqrt{12}$. The location of the center of the target window is described by its offset from the nominal design value (in this example, this value is 6 μm). This offset of the center of the target window is denoted as $T_{offset}$. This value is also in the unit of $\sigma_k$, and it shows how far away the calibration target can be from the nominal design value.

We first study the trade-off between calibration success rate and calibration target window size when the calibration target is right at the nominal design value ($T_{offset} = 0$). For the ESES method, a set of different values for common difference of the arithmetic sequence are analyzed, which are $d_{ESES} = \sigma_k/4$, $\sigma_k/2$, $\sigma_k$ and $2\sigma_k$. The target window size $T_{window}$ varies from $\sigma_k/100$ to $\sigma_k/5$ in this experiment. For each different SES/ESES setting and target window size, we run the Monte Carlo simulation with $10^5$ samples. Figure 3.3 shows the relationship between calibration failure rate and calibration target window size. In order to have a closer look at the region where success rate is very close to 100%, failure rate in log scale are shown in this figure.

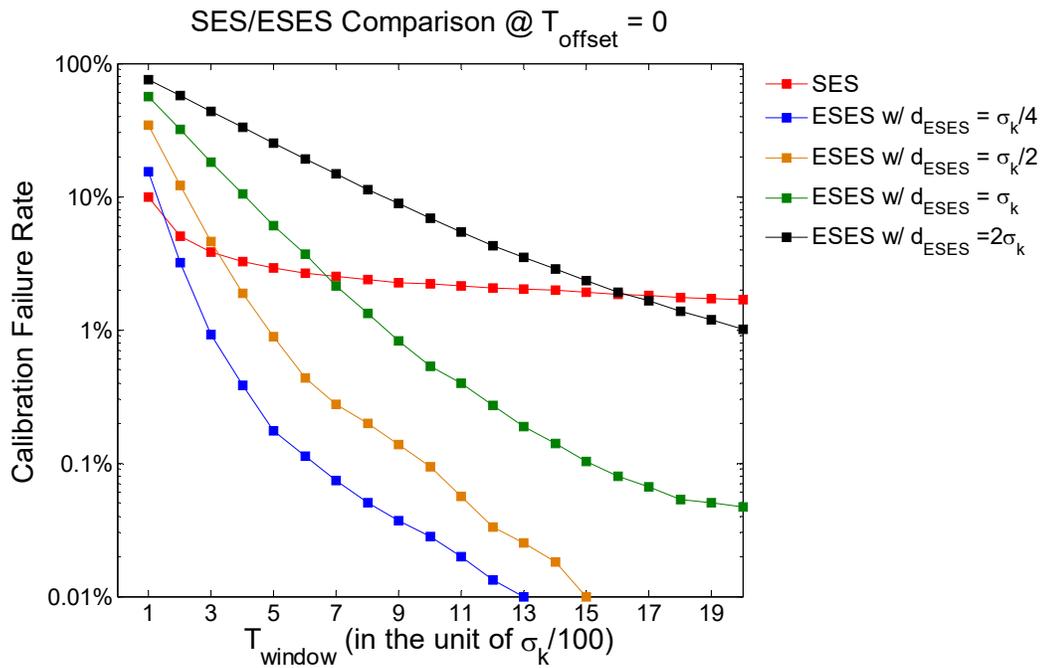

Figure 3.3 SES/ESES calibration failure rate vs. calibration target window when $T_{offset} = 0$.

From Figure 3.3, we can see that the SES method only shows advantage over the ESES method in terms of lower calibration failure rate when the calibration target window size $T_{window}$ is set as minimum. As $T_{window}$ increases the failure rate of the ESES method improves much faster than that of the SES method improves. Especially at the region where calibration failure rate is lower than 10% as shown in Figure 3.3, the ESES method is dominating the SES method for providing higher calibration yield. The inefficiency of the SES method when the calibration target is right at the nominal design value can be explained as follows: although SES provides more design choices than ESES does at the center of the design choice distributions as shown in Figure 3.2 (a) and Figure 3.2 (b), which supposedly translate to higher calibration yield for the SES method at that region, the center of the design choices distribution is not the calibration target for $T_{offset} = 0$. The calibration target is always a constant. And in the case of $T_{offset} = 0$, the calibration target is the nominal design value, which is exactly 6 μm in this example. Meanwhile the center of the design choices distribution is a random variable. It is determined by the sizes of the N elements that are available in each sample. Therefore, statistically, the center of the design choices distribution of the SES method can be far away from the nominal design value. And because SES has a very sharp distribution, as shown in Figure 3.2 (a), the misalignment between the center of the design-choices distribution and the nominal design value can result in a very small number of design choices at the nominal design value. Meanwhile for the ESES method, as the design choices have a much wider distribution, the misalignment between the center of the design choices and the nominal design value has less impact on the design choices density at the nominal design value. Therefore, although SES has more design choices around the center of the design choices distribution, it does not translate to a higher calibration yield as compared to ESES when the calibration target is the nominal design value.

As the calibration target location deviates from the nominal design value as $T_{offset}$ increases, the ESES design method is showing increasing advantages over the SES method. Figure 3.4 shows the calibration failure rate vs. $T_{window}$ for $T_{offset} = \sigma_k$ in solid lines, while dotted lines show the results for $T_{offset} = 0$ for comparison. The failure rate for $T_{offset} = \sigma_k$ increases dramatically for the SES method for almost one order of magnitude as compared to the case of $T_{offset} = 0$. Meanwhile for the ESES method, the calibration failure rate only increases marginally as $T_{offset}$ increases to $\sigma_k$.

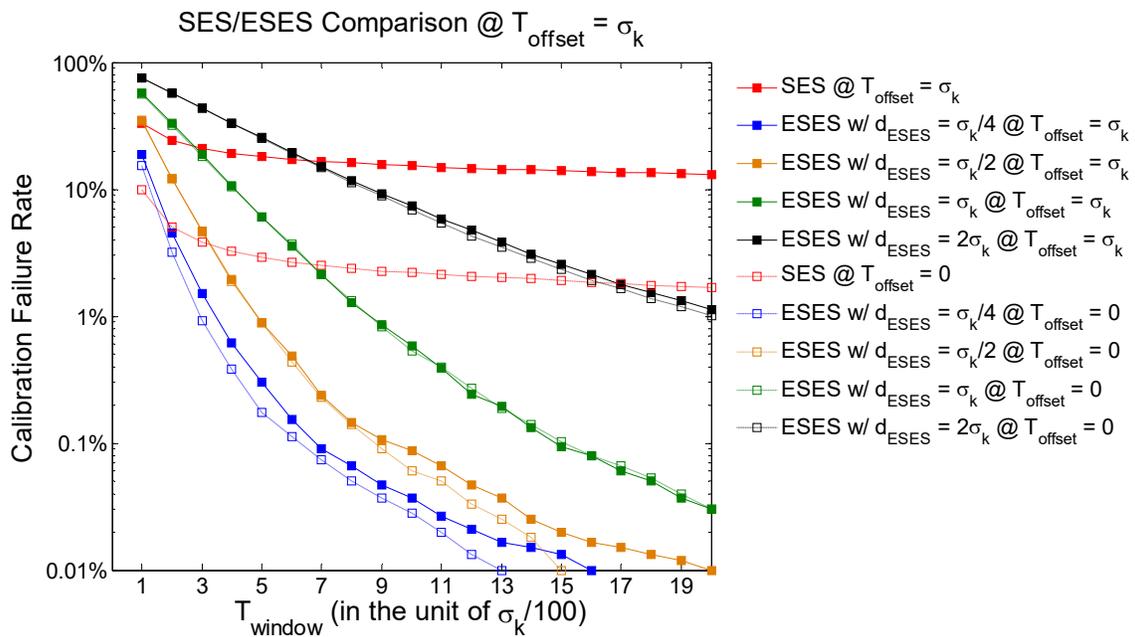

Figure 3.4 SES/ESES calibration failure rate vs. calibration target window when $T_{offset} = \sigma_k$.

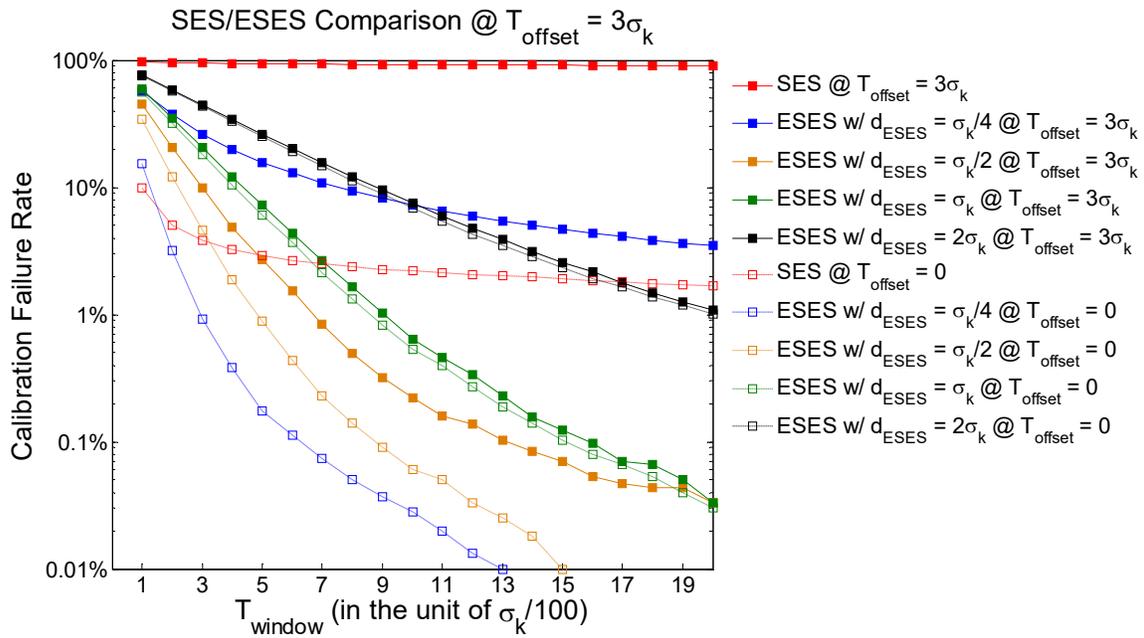

Figure 3.5 SES/ESES calibration failure rate vs. calibration target window when $T_{offset} = 3\sigma_k$.

As $T_{offset}$ keeps increasing, we very soon reach a point that the SES method is no longer usable for calibration while the ESES method is still effective. Figure 3.5 shows the calibration failure rate vs. $T_{window}$ for $T_{offset} = 3\sigma_k$ in solid lines, while dotted lines show the results for $T_{offset} = 0$ for comparison. For the SES method, the calibration failure rate is almost 100% when $T_{offset} = 3\sigma_k$. For the ESES method, the calibration failure rate for the case of $d_{ESES} = \sigma_k/4$ also increases significantly, which means the calibration range of this specific ESES setting is not sufficient for $T_{offset} = 3\sigma_k$. But as $d_{ESES}$ increases, the ESES method can still provide very low calibration failure rate. For example, when $d_{ESES} = \sigma_k/2$ and $T_{window} = 7\sigma_k/100$, the calibration failure rate is down below 1%.

The aforementioned experiments are all performed at fixed $T_{offset}$ values. In a more realistic calibration environment, the deviation of the calibration target from the nominal design value is determined by the random variations of other variation sources in the design. Therefore, in the next experiment we assume $T_{offset}$ is a random number that follows a normal distribution $N(0, \sigma_T)$, where $\sigma_T$ shows the total variation coming from the other variation sources in the design. Before calibration, overall standard deviation in the design, denoted as $\sigma_{All}$, can be calculated as $\sigma_{All} = \sqrt{(\sigma_k^2 + \sigma_T^2)}$. As aforementioned, after successfully performing calibration, the standard deviation can be reduced to $T_{window}/\sqrt{12}$ as the calibrated values are all bounded in a small calibration window. To show the benefit of the calibration, we further define the ratio of $\sigma_{All}$ and $T_{window}/\sqrt{12}$, denoted as $R_{cal}$, as the standard deviation reduction factor of the calibration method. For the following experiments, $\sigma_T$ is set as $\sigma_k/4$, $\sigma_k$ and $2\sigma_k$. These three values correspond to three scenarios: other variation sources in the design are negligible, comparable or dominant as compared to the variations of the N-element undergoing SES/ESES based calibration process. Figure 3.6 to Figure 3.8 show the calibration failure rate vs. calibration target window size for these three scenarios respectively.

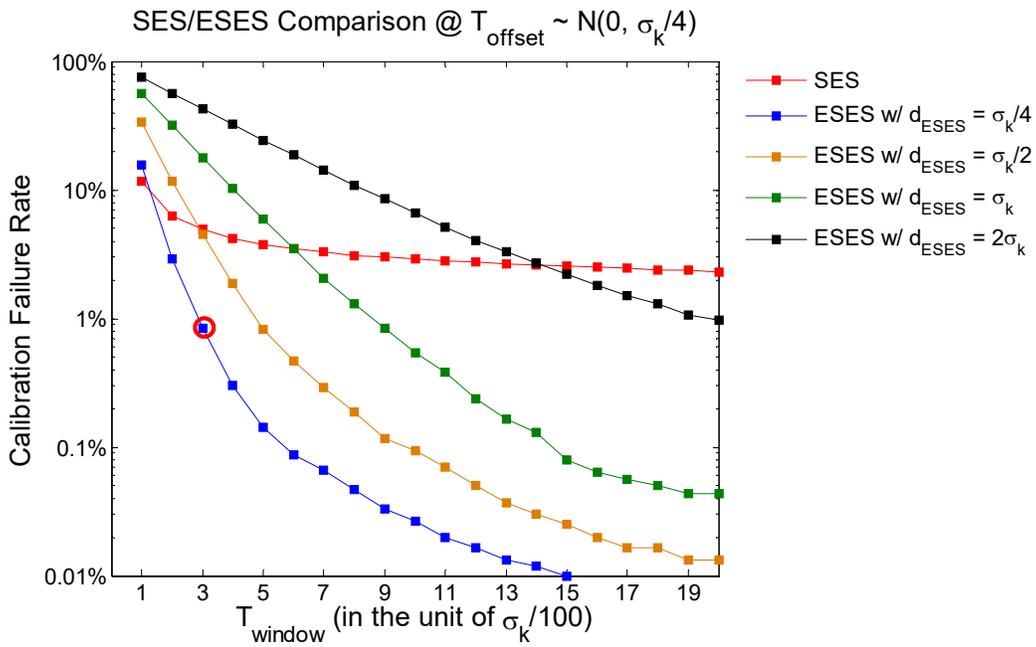

Figure 3.6 SES/ESES calibration failure rate vs. calibration target window when $T_{offset} \sim N(0, \sigma_k/4)$.

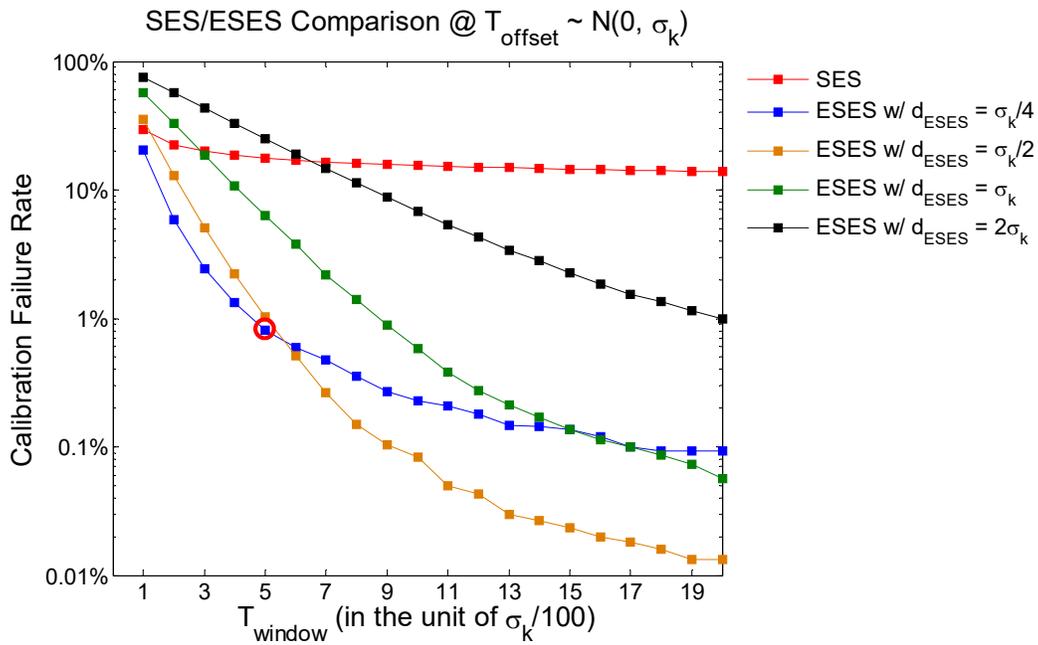

Figure 3.7 SES/ESES calibration failure rate vs. calibration target window when $T_{offset} \sim N(0, \sigma_k)$.

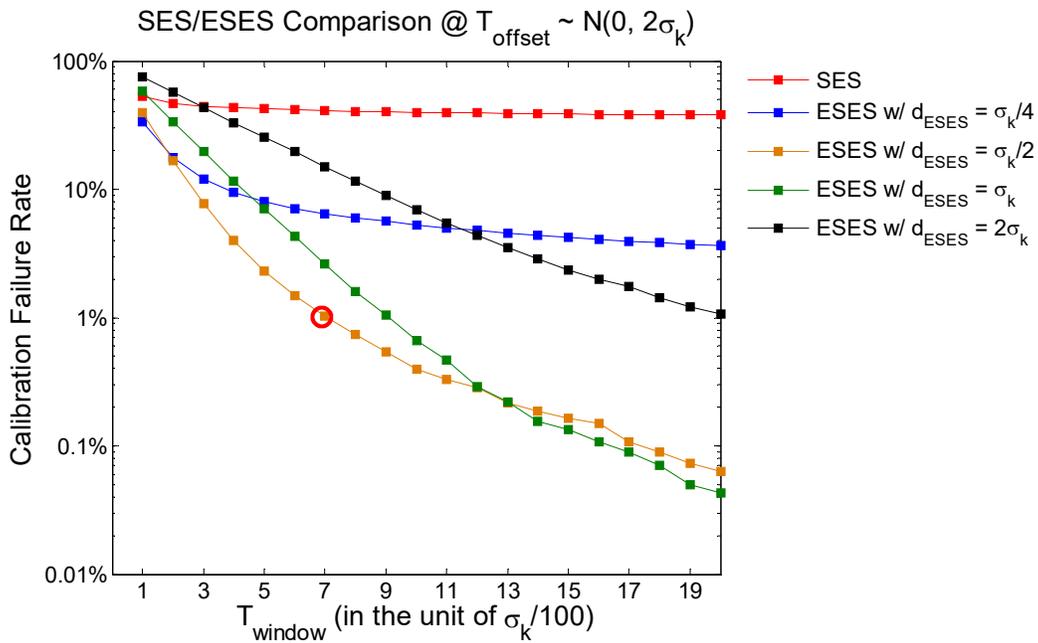

Figure 3.8 SES/ESES calibration failure rate vs. calibration target window when $T_{offset} \sim N(0, 2\sigma_k)$.

In Figure 3.6, where the other variation sources in the design are negligible, the calibration failure rate of the SES method is much higher than 1%. However, for the ESES method, calibration failure rate can be much lower than 1% depending on the value of calibration target window. For example, for $T_{window} = 3\sigma_k/100$, which corresponds to a standard deviation reduction factor $R_{cal}$ of 115, the calibration failure rate is less than 1% with $d_{ESES} = \sigma_k/4$ for ESES method.

In Figure 3.7 and Figure 3.8, where $\sigma_T$ increases to $\sigma_k$ and $2\sigma_k$ such that other variation sources become comparable or dominant to the N-element, the calibration failure rate of the SES method becomes unacceptable, as it is much higher than 10%. On the other hand, the ESES method still exhibits very low calibration failure rate for certain calibration windows. For example, if we still target at a calibration failure rate less than 1%, for $\sigma_T = \sigma_k$ (Figure 3.7), we can obtain a $R_{cal}$ of 98 when $d_{ESES} = \sigma_k/4$ and $T_{window} = 5\sigma_k/100$; for $\sigma_T = 2\sigma_k$ (Figure 3.8), we can obtain a $R_{cal}$ of 111 when $d_{ESES} = \sigma_k/2$ and $T_{window} = 7\sigma_k/100$.

The above experiments show that the ESES method is effective to perform calibration with more than 99% calibration success rate while achieving a standard deviation reduction factor of around 100 in all three scenarios. This also means that the matching property after calibration is improved by approximately 40 dB.

While $R_{cal}$ shows the matching improvement the ESES method can provide, $\sigma_T$ (in the unit of $\sigma_k$) shows the amount of variations from other variation sources that the ESES method can handle. A higher $\sigma_T$ value means the ESES method needs to provide a larger calibration range to cover the overall variations. In the following experiment, we further study how much matching improvement we can achieve while we keep increasing $\sigma_T$. Assuming our calibration yield target is greater than 99%, Figure 3.9 shows the relationship between standard deviation reduction factor $R_{cal}$ and $\sigma_T$ (in the unit of $\sigma_k$). As we can see from the figure, different $d_{ESES}$ settings are needed to achieve optimal $R_{cal}$ at different $\sigma_T$ values. For example, the ESES method can achieve an $R_{cal}$ value around 100 when $\sigma_T$ is no greater than $9\sigma_k$. As $\sigma_T$ keeps increasing, the achievable $R_{cal}$ value decreases. In order to obtain an $R_{cal}$ value greater than 50, $\sigma_T$ can be as large as $15\sigma_k$. In both aforementioned $R_{cal}$ targets, the allowable $\sigma_T$ value can be almost one order of magnitude larger than $\sigma_k$. These numbers show that even the variations from other sources are much larger than those of the N-element, we can achieve very large matching property improvement by using the ESES method which leads to high-resolution calibration results.

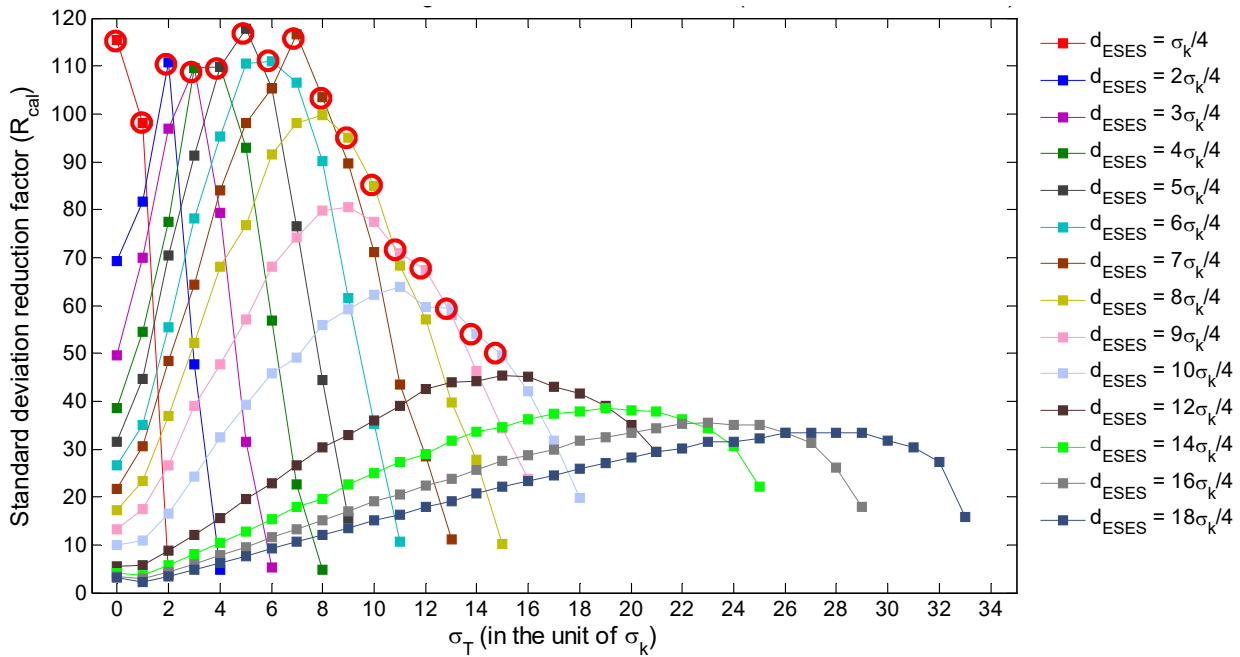

Figure 3.9 ESES calibration standard deviation reduction factor $R_{cal}$ vs. $\sigma_T$ with more than 99% calibration success rate.

For the ESES method, which uses arithmetic sequence for element sizing, all of the previous simulation results are obtained by setting the average value of the sequence (denoted as $a_{ESES}$) to be 1 μm. We also assumed that the standard deviation of each element is proportional to the square root of its nominal size, and then set the center value of the standard deviation for the distribution of the 12 elements to be 0.01 μm. This results in a relative standard deviation for each element to be about 1% before calibration, which is considered as a very good starting point for matching. If we have a worse starting point for the relative standard deviation, it can be shown that the simulation results obtained previously still withstand. For example, we now set the $a_{ESES}$ value to be 1μm, 0.5 μm, 0.25 μm, 0.125 μm or 0.0625 μm, while keeping the center value of the standard deviations of the 12 elements fixed at 0.01 μm. As a result, the relative standard deviation of each element is approximately 1%, 2%, 4%, 8% and 16% respectively before calibration. We also set $d_{ESES} = \sigma_k/4$ for all $a_{ESES}$ cases. Now, for the extreme case when $a_{ESES} = 0.0625$ μm, the nominal sizes for the 12 elements are of an arithmetic sequence starting from 0.0288 μm to 0.0962 μm, with a common difference of 0.0061 μm. Whereas for $a_{ESES} = 1$ μm, the nominal sizes range from 0.9663 μm to 1.0337 μm for the 12 elements, while having a same common difference of 0.0061 μm for the arithmetic sequence. The simulation results for the calibration failure rates when $T_{offset} = 0$ are shown in Figure 3.10 for different $a_{ESES}$ settings.

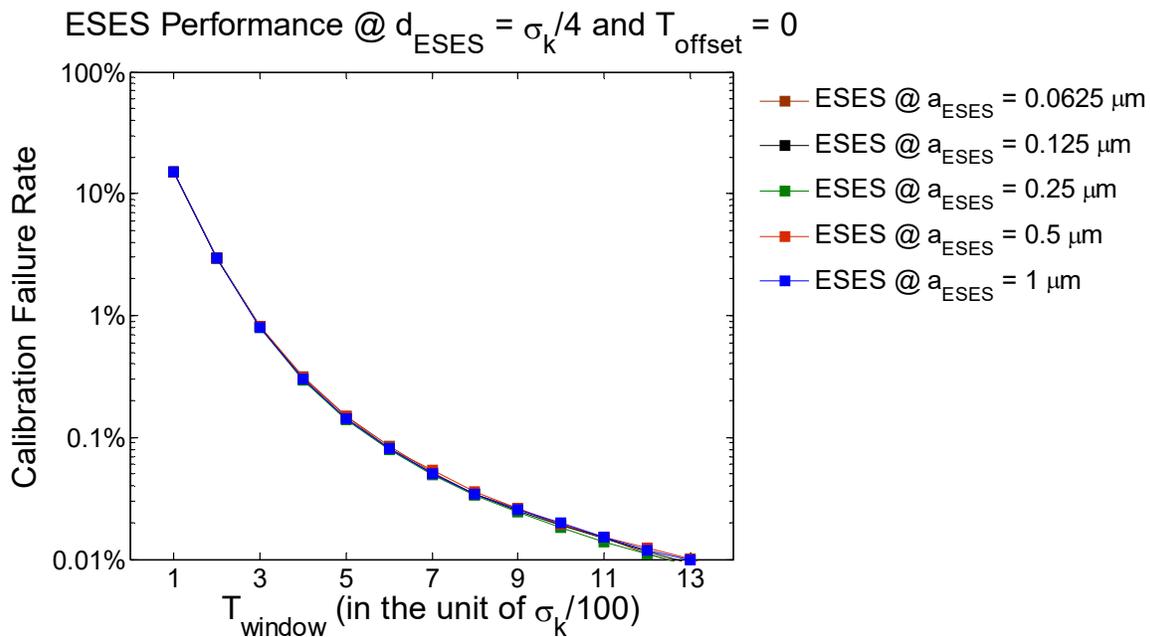

Figure 3.10 ESES calibration failure rate for various $a_{ESES}$ settings when $T_{offset} = 0$, $d_{ESES} = \sigma_k/4$

As observed from Figure 3.10, the simulation results for different $a_{ESES}$ values almost overlap with each other. Similar behaviors can be observed for different $T_{offset}$ settings and different $d_{ESES}$ settings. This suggests that the results obtained in this section do not lose generality because of the specific relative standard deviation settings (around 1% for each ESES-based element) that we have used in the aforementioned example.

To summarize the results of the experiments in this section, by using the proposed ESES method we can utilize combinatorial redundancy in a non-dominant variation source to calibrate the dominant variations in the design and still achieve large improvement for the matching property. This proposed calibration method extends the usage of combinatorial redundancy for general high-resolution calibration applications. As the calibration location no longer has to be the dominant variation source, it provides the flexibility of choosing a suitable location in the design for applying combinatorial redundancy. Two types of applications, current calibration and phase delay calibration, are considered in the following sections.

## 3.2 ESES-based High Resolution Current Calibration

The combinatorial redundancy based ESES method can be applied to current calibration to achieve high calibration resolution with more than one order of magnitude matching property improvement. Consider the conceptual circuit shown in Figure 3.11. The current source under calibration is split into N non-uniformly sized sub-current sources, with each one controlled by a switch and only K of the N sub-current sources are activated. During calibration, different K subsets can be tested in order to find the optimal selection from the built-in combinatorial redundancy. The number of available combinations increases exponentially as N and K increase, providing a large number of calibration choices and enabling high calibration resolution. The applied ESES method provides higher calibration range as compared to SES method to accommodate extra variation sources if any. Even in the case of there is no extra variation source of the current source in the design, ESES can provide higher calibration success rate as shown in Figure 3.3.

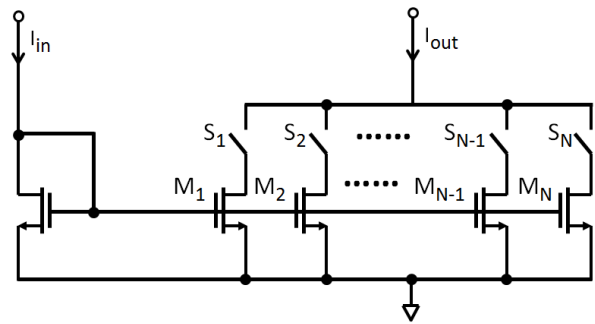

Figure 3.11 Conceptual circuit of ESES-based current calibration.

In contrast to traditional current calibration, the ESES-based high resolution current calibration method does not involve the overhead of a CALDAC, as seen in other calibration methods. The major overhead of this method is the sacrificed circuit area of N-K unselected sub-current sources. However, as pointed out in [3], the combinatorial redundancy dramatically relaxes the area requirement for achieving certain CMOS matching properties. As a result, even considering the sacrificial area, the total area for current source can be reduced significantly (as compared with traditional sizing) when meeting the high matching requirement.

There are various applications for the ESES-based high resolution current calibration method. One direct application for current source calibration is segmented CS-DACs, where the current matching for the thermometer coded several bits MSBs is critical for achieving good D/A linearity. The high resolution current calibration method can be applied to the MSBs unary current sources to optimize matching properties. This application is shown in the design of a ESES-based CS-DAC in section 5.3.1.

Another application is for differential amplifier transconductance calibration. A conceptual circuit is shown in Figure 3.12. The tail current source of the differential amplifier is split into several sub-current sources to create combinatorial redundancy. As the transconductance of the differential pair is set by the DC biasing current, by tuning the biasing current, the transconductance is effectively calibrated. This application is demonstrated in the design of a transconductance amplifier that is described in section 4.3.2.

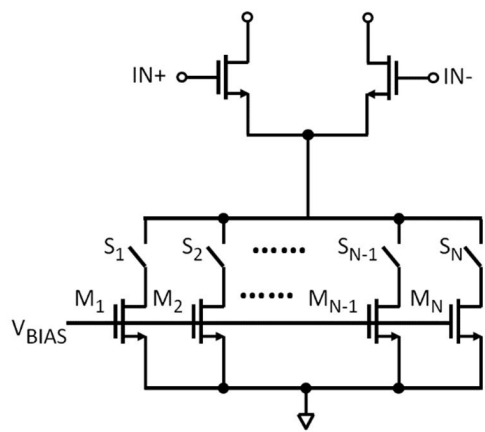

Figure 3.12 Conceptual circuit of ESES-based differential amplifier transconductance calibration.

## 3.3 ESES-based High Resolution Phase Delay Calibration

The combinatorial redundancy can further be applied to phase delay calibration for high calibration resolution and the conceptual circuit is shown in Figure 3.13 (a). The idea is to break the NMOS and PMOS transistors into N elements, and select a subset of them by adding big switches at their drain nodes (to minimize "on" resistance of the switch). The ESES design methodology with wider tuning range is suitable for phase delay calibration since a chain of logic gates is usually involved in the design and the delay variation generally not only comes from a single stage. By using the ESES methodology, calibration performed in one inverter stage can potentially cover the delay variations coming from the entire logic chain.

Different from existing phase delay calibration methods, the circuit shown in Figure 3.13 (a) has the capability of tuning the rising edge and falling edge of the inverter output independently. If only one edge is needed for calibration, e.g. falling edge, the circuit can be simplified, as shown in Figure 3.13 (b).

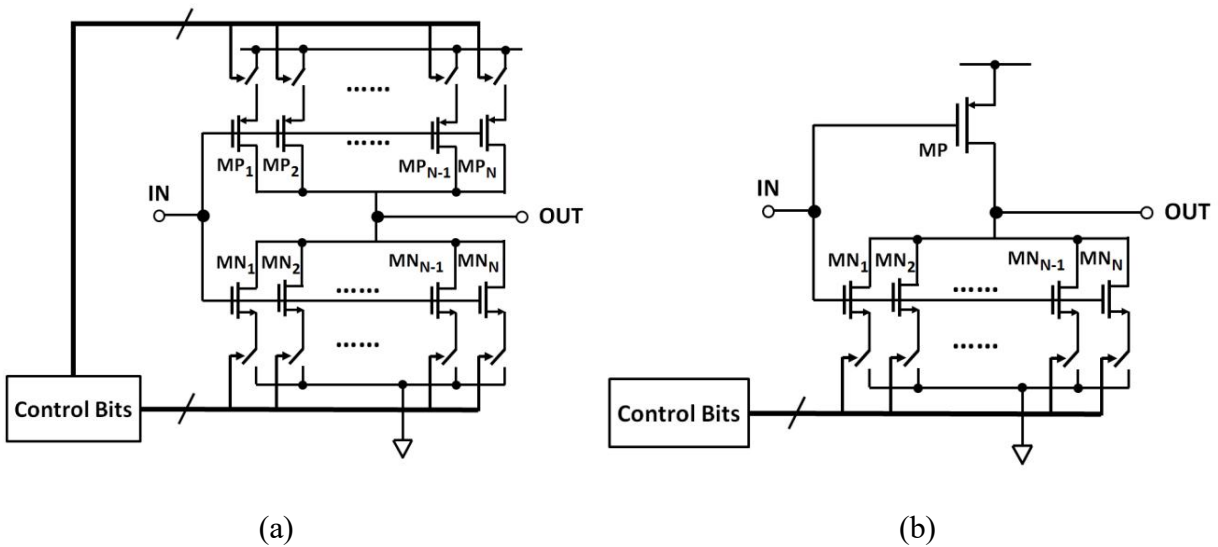

Figure 3.13 Conceptual circuit of ESES-based phase delay calibration:
(a) tunable rising and falling edge; (a) tunable falling edge only.

This high resolution phase delay calibration can be applied to various circuit designs. The delay of asynchronous signals can be calibrated directly using this calibration technique. One application for analog/RF design is the calibration of a pair of differential clocks. As shown in Figure 3.14 (a), a pair of differential clocks CLK_IN_P and CLK_IN_N pass through a pair of tunable inverters (Figure 3.13 (a)) and one more pair of inverters for buffering. By calibrating both the rising edges and falling edges, the even-order harmonics for the differential output (CLK_OUT_P – CLK_OUT_N) can be minimized. This can improve even-order distortions for analog/RF design driven by the differential clock pair. One implementation of this application is shown in section 4.3.5.

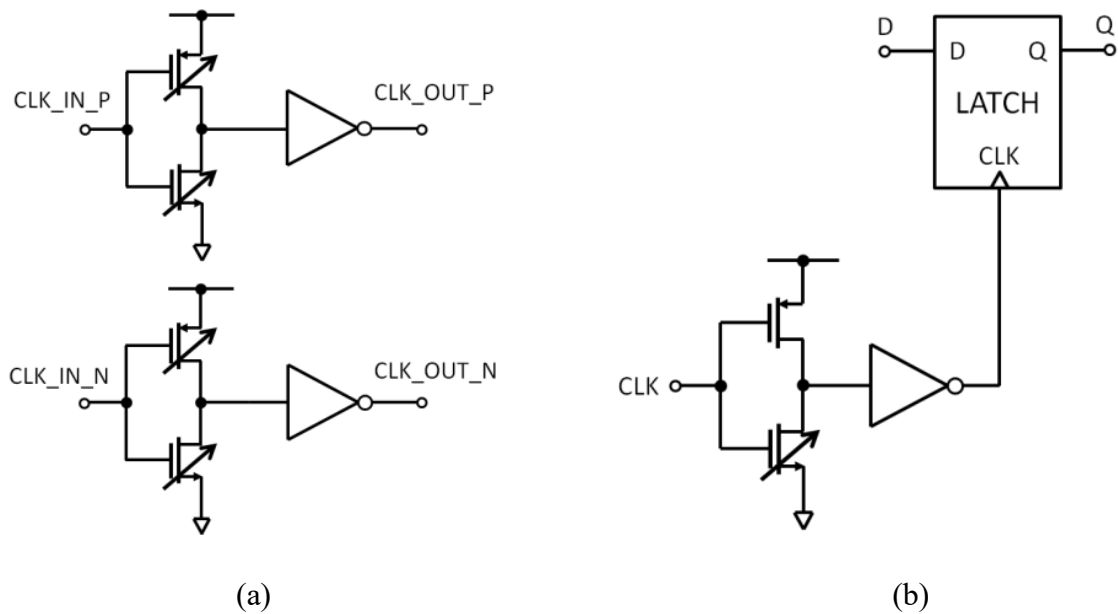

(a)            (b)

Figure 3.14 Conceptual circuits for high resolution phase delay calibration applications: (a) differential signals calibration; (b) delay calibration of synchronous signals.

Many digital signals used in analog/RF design have to be synchronous, e.g. switching signals for D/A converter and multi-phase LO signals generated by a frequency divider. For those signals, the delay calibration can be done indirectly on the clocking signals. Figure 3.14 (b) shows the application of delay calibration for synchronized signal Q. Assume that the latch is transparent when clock is high and there is no switching activity of D when clock is high. The clock signal is passed through a tunable inverter (Figure 3.13 (b)) to calibrate the rising edge for the latch's clock input, hence calibrating the transitioning timing of output signal Q. The benefit of calibrating over clocking signals instead of output Q signal directly is that the former one applies to all signals generated by this latch. This is especially useful for analog/RF design where differential latches with differential outputs are widely used. One implementation of delay calibration for synchronous signals is presented in section 4.3.5 for the design of a multi-phase LO generator circuit. A second implementation is shown in section 5.3.2 for the design of a differential latch with delay and duty-cycle calibration.

## 3.4 Summary

In this chapter, a new design method call extended statistical element selection (ESES) was introduced. The ESES method can provide wider calibration range as well as higher calibration yield as compared to the original SES method. The new method also enables the flexibility of the calibration location in a design. With these advantages, the ESES method can be applied to a broader range of applications as compared to SES while still providing high calibration resolution. Two types of ESES-based calibration were also introduced in this chapter. One is current source calibration, and the other is phase delay calibration. Applications for both types are further discussed. In the next two following chapters, we are going to present two circuit designs that are utilizing ESES-based calibration methods.

# 4 Harmonic Rejection Receiver Design

The harmonic rejection scheme presented in [28] can be applied to a wideband receiver design for rejecting interferences at local oscillator's harmonic frequencies. However, the achieved harmonic-rejection ratio highly depends on the gain matching and phase matching. As a demonstration of the proposed ESES-based high resolution calibration method, we present a design of a harmonic-rejection receiver that utilizes ESES for gain and phase calibration that achieves best-in-class harmonic-rejection ratios [29] [30].

## 4.1 Harmonic Rejection Receiver Overview

### 4.1.1 Harmonic Rejection Scheme and Impacts of Gain and Phase Errors

Due to the proliferation of wireless devices and standards, software defined radio (SDR) has become an increasingly interesting research area for radio architectures, not only to replace multiple narrow-band radios with reduced size and cost, but also because SDR is an important enabler for future cognitive radios with dynamic, intelligent, and thus efficient spectrum usage.

On the receiver side, SDR needs to support the reception of signals in multiple frequency bands with different radio standards. Thus a wideband receiver seems to be an obvious solution for a SDR receiver [31]. However, wideband receivers not only receive the desired signal, but also interfering signals, which can ultimately degrade receiver performance. In particular, interfering signals at harmonics of the local oscillator (LO) are especially problematic. Figure 4.1 illustrates the reception of interference at LO harmonic frequencies. Harmonics of the LO downconvert the interference to the same frequency as the desired signal thereby reducing the signal-to-noise ratio. The interfering signals need to be rejected down to thermal noise levels to minimize the desensitization of the receiver. This rejection is typically performed with a band-select filter but this solution is not ideal for SDRs. In [31] it was shown that a -40 dBm to 0 dBm interferer at the harmonics of the LO requires a rejection ratio of 60 dB to 100 dB.

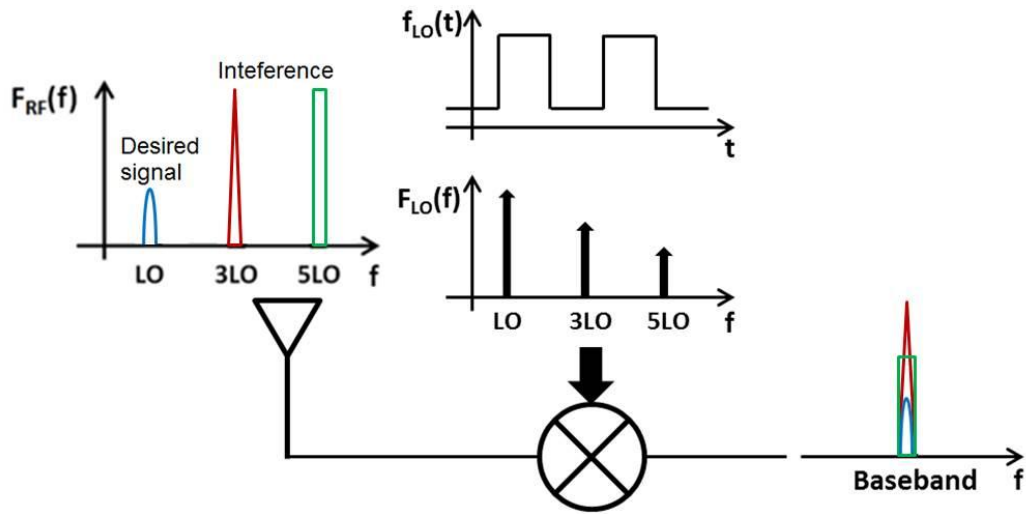

Figure 4.1 Reception of interference at LO harmonic frequencies for wideband receiver.

A harmonic-rejection mixer scheme, which was first proposed in [28] for a transmitter design, has been widely adopted in modern wideband RF receiver designs [32] [33] [34] [35] [36] [37] [38] [39] to alleviate the wideband interference problem. The basic idea is to construct an equivalent LO signal that is free of $3^{rd}$ and $5^{th}$ order harmonics, such that the $3^{rd}$ and $5^{th}$ order harmonic interferer would not be down-converted by the mixer, therefore being completely rejected. To achieve this, three LO signals need to be generated with a 45° phase shift from one to another and summed up by a weighting ratio of 1: $\sqrt{2}$ : 1. However, the harmonic rejection effect degrades as the phase shift deviates from 45° and gain ratio deviates from $\sqrt{2}$. Practically, with gain errors and phase errors considered in that are due to CMOS technology process variation as well as difficulties of generating the irrational number of $\sqrt{2}$, the achievable harmonic rejection ratio (HRR) is limited to 30-40 dB [32]. Figure 4.2 illustrates that a harmonic rejection wideband receiver still has limited harmonic rejection effects if gain errors and phase errors exist.

Other than 3$^{rd}$ and 5$^{th}$ order harmonic interference, wideband receivers also need to reject even-order harmonic interference, specifically 2$^{nd}$, 4$^{th}$ and 6$^{th}$ order harmonics, which are still not far enough away from the desired signal band to allow for simple filtering. Fortunately, the differential nature of the mixer operation makes the wideband receiver ideally immune from even-order harmonic interference. If the mixer has perfect differential phase, meaning exactly180° of phase shift between the differential signals, the even-order harmonic rejection ratio is infinite. However, in reality the phase deviates from 180° resulting in degraded even-order HR performance. In [31], the second order HRR (HRR$_2$) was reported as 62 dB at 0.8 GHz LO frequency. At higher frequency bands, the effect of phase mismatch would be more critical, thereby resulting in even worse HRR performance.

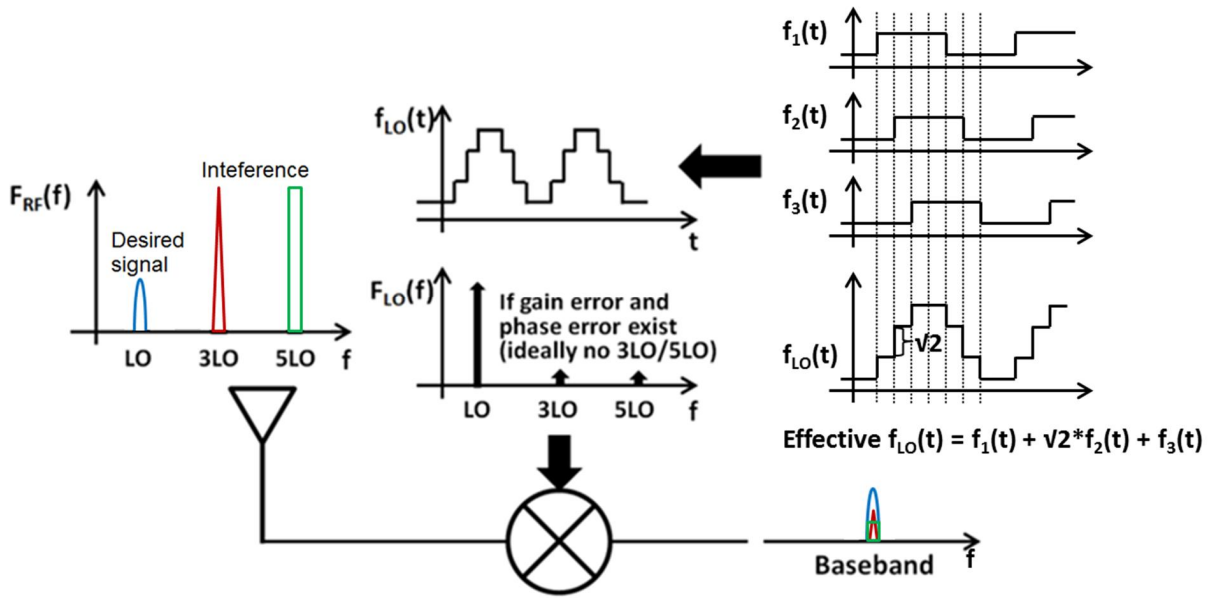

Figure 4.2 Reception of interference at LO harmonic frequencies for harmonic rejection wideband receiver with gain errors and phase errors.

### 4.1.2 Previous Work on Achieving High HRR

Generating the irrational number of √2 accurately has been a very difficult task to do on chip for achieving high 3$^{rd}$ and 5$^{th}$ order HRR (HRR$_3$, HRR$_5$). It is usually done through approximating √2 by 3/2, 7/5, 17/12, or 41/29, with increasing HRR performance in this order. The maximum achievable HRR$_{3,5}$ with a 41/29 approximation is more than 77 dB. However, considering random variations in CMOS processes, this theoretical number is difficult to achieve. To solve the gain mismatch problem of the HR mixer, a two-stage HR scheme was proposed in [33], and has been widely adopted recently [34] [35] [40]. The fundamental idea is to perform HR operations twice in baseband, such that the combined HRR is the summation of the HRR of the two stages. The authors also proved in [31] that the overall gain error is the multiplication of the gain errors of the two stages. Therefore, a very small overall gain error can be achieved. The authors successfully approximate 41/29 with negligible gain errors in [33], which can potentially reach an HRR of more than 77 dB. While this approach accounts for gain mismatch, it cannot improve the errors coming from LO phase mismatch. As the first stage and second stage suffer from the same LO phase error, the correlated phase errors in two stages do not benefit from this two-stage operation. Therefore, LO phase mismatch limits the achievable HRR of the two-stage HR to approximately 60 dB without extra RF filtering [33] [34] [35] [40]. Furthermore, this phase mismatch increases with RF frequency, thus further limiting the usage of the two-stage HR scheme in higher frequency bands. It is also worth mentioning that this method requires extra baseband circuitry to perform the second-stage HR operation.

In addition to relying on the intrinsic matching of CMOS process, calibration methods [36] [37] have also been proposed to increase the HRR. These methods rely on a common technique for which gain mismatch cancels phase mismatch. However, different phase and gain setting are needed to optimize HRR$_3$ and HRR$_5$, respectively, resulting in the optimization of either HRR$_3$ or HRR$_5$, not both. In [36] and [37] the authors calibrate the gain mismatch of the HR mixers and achieve HRR$_3$ of approximately 70 dB. However, a degradation of HRR$_5$ after HRR$_3$ calibration is reported in [36]. In [38], both gain and phase are tunable and the achieved HRR$_3$ is limited to 65 dB. Although these techniques cannot optimize HRR$_3$ and HRR$_5$ concurrently, they are still useful when the system is able to detect a single harmonic jammer and re-configure the calibration.

Other than these aforementioned traditional circuit calibration techniques, a unique HR scheme with adaptive interference cancellation was proposed in [39]. This scheme dynamically estimates and then subtracts harmonic interference at the baseband in digital domain. To do so requires a total of four baseband ADCs, which is the major circuit overhead for this scheme. Although the reported HRR can be more than 80 dB, as the technique can only estimate one interference signal, it ends up either rejecting 3$^{rd}$ order harmonic or 5$^{th}$ order harmonic interference.

In summary, the previous techniques have all failed to achieve simultaneous HRR$_3$ and HRR$_5$ above 70 dB. In addition, as mentioned in section in 4.1.1, even-order harmonic rejection, specifically 2$^{nd}$ order, 4$^{th}$ order and 6$^{th}$ order HRR (HRR$_2$, HRR$_4$, and HRR$_6$) are also important for receiver performance and need to be taken care of to achieve high HRR values. Not many previous works have addressed this problem. In [38] a duty-cycle calibration is performed to calibrate HRR$_2$ and the resulting even-order rejection is reported to be greater than 65 dB.

## 4.2 ESES-based Harmonic Rejection Receiver Design

To overcome the aforementioned shortcomings of the previous solutions, we are aming to design an HR receiver that can achieve high HRR values for $2^{nd}$ to $6^{th}$ order harmonic interference simultaneously and high HRR values for high frequency bands. As an example of the ESES-based calibration method we applied it for the HR receiver calibration. To achieve better HRR, a digital calibration scheme is used to tune both gain mismatch and phase mismatch for $HRR_{3,5}$. Gain mismatch and phase mismatch can be decoupled and calibrated independently following ESES method. Phase mismatch for $HRR_{2,4,6}$ can also be calibrated by ESES method. With the calibration accuracy provided by ESES approach, better $HRR_{2-6}$ can be expected at not only low frequency band but also high frequency band.

Our proposed HR receiver architecture is shown in Figure 4.3. For simplicity, all paths are drawn as single-ended but in the actual implementation only the RF input is single-ended while the rest of the receiver is fully differential. The receiver frontend starts with an LNA, which performs single-ended to differential conversion while achieving low noise figure using noise-cancellation technique [41]. The LNA is followed by a transconductance ($G_m$) stage, which is a fully differential design for better second order in-band linearity [42]. The Gm stage is followed by a standard passive mixer and baseband transimpedance amplifier (TIA) hereafter. The TIA is realized as an operational transconductance amplifier (OTA) configured in closed loop. For HR operation, four branches are created after the LNA and they are followed by a harmonic recombination stage for each respective I and Q path. A switch preceding each TIA is used to turn each branch on or off, which is needed in the ESES calibration process.

Gain mismatch calibration in the HR receiver is achieved by tuning the gain of the Gm stage with the ESES method, as described in the upcoming section 4.3.2. The harmonic recombination stage for both I and Q paths is designed as weighted resistors and a summing TIA stage. The weighting ratio for the 0°, 45° and 90° branches is 12:17:12. The inherent limitation of this weighting ratio is also accounted for by using the ESES method.

The phases of the multi-phase LO generator are designed to be tunable by utilizing the ESES method for calibrating various phase mismatches in the HR receiver. This is covered in the upcoming section 4.3.5.

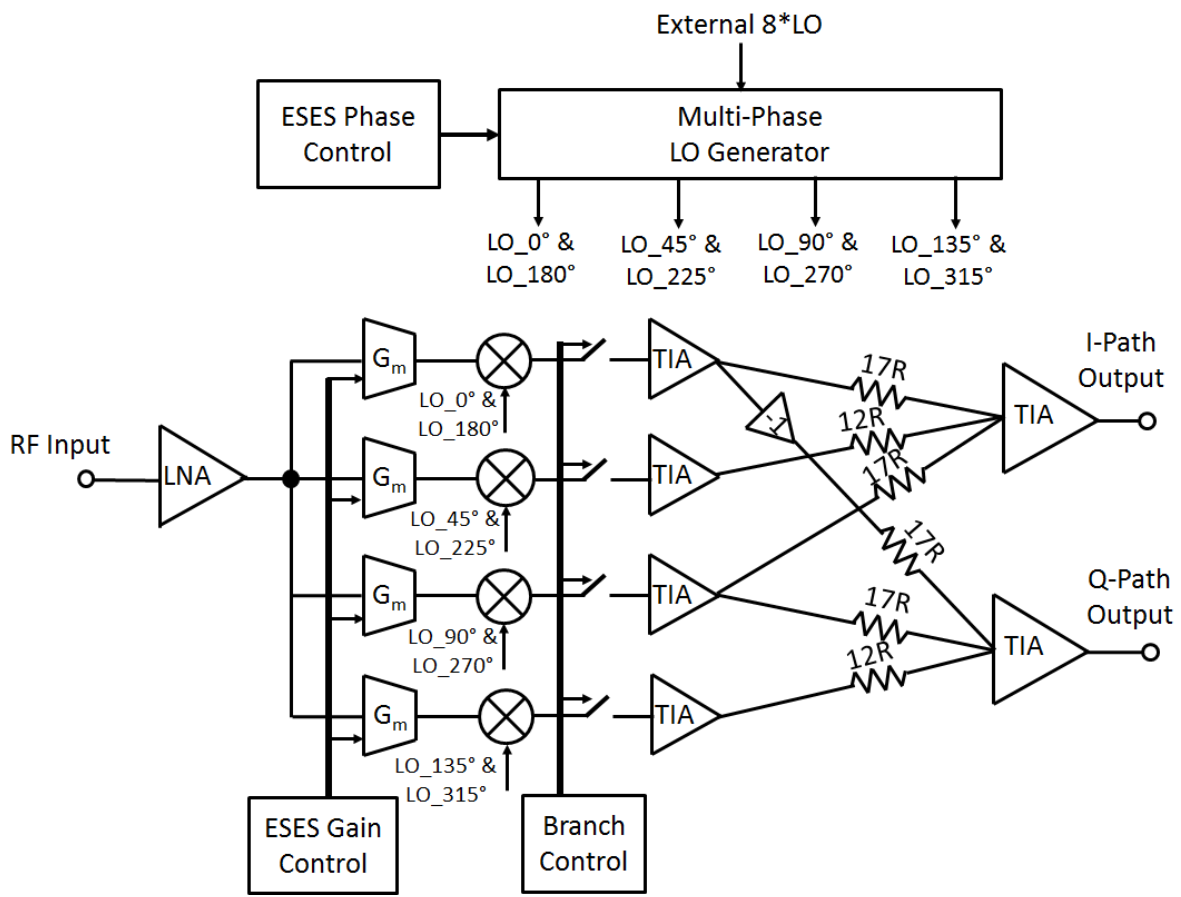

Figure 4.3 HR receiver architecture with ESES-based calibration.

## 4.3 Chip Implementation

### 4.3.1 LNA Design

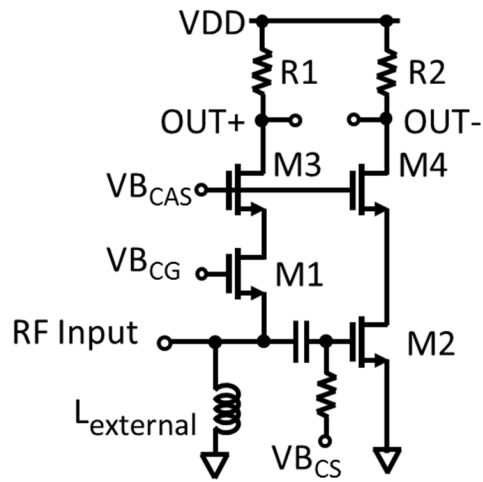

Figure 4.4 LNA circuit design.

The LNA circuit design is shown in Figure 4.4. A similar topology was proposed in [41] for achieving simultaneous output balancing, noise-cancellation and distortion-cancellation. The common gate transistor M1 was designed such that the real part of the LNA input impedance, which is roughly $1/g_{m1}$, is matched to the 50 Ω source impedance. The noise or distortion induced by the input matching transistor M1 has out-of-phase responses at the RF input node and positive output node. A common source transistor M2 was then added such that the noise or distortion at the RF input node further creates an out-of-phase response at the negative output node. Therefore, the noise or distortion coming from M1 has in-phase responses at the positive and negative LNA output which results in cancellation for differential operation. The condition for M1 noise and/or distortion cancellation as in [41] is $g_{m2}/g_{m1} = R_1/R_2$ whereas $1/g_{m1} = 50$ Ω. On the signal side, the common gate stage has a positive gain while the common source stage has negative gain. The differential output voltage gain is the summation of the gains of both branches, which are both $R_1/50\Omega$ when simultaneously achieving both power matching and noise cancellation. In addition to the noise cancellation, this LNA operates as a balun for converting the single-ended RF input to a differential RF output for succeeding stages.

In this design, $R_1$ was set to 240 $\Omega$, which results in a gain of 19.6 dB for the LNA in simulation. The transconductance of M1 was designed as 20 mA/V to meet the receiver power matching requirement. For M2, the transconductance was designed to be 80 mA/V as a trade-off between power consumption and noise figure (NF). Power consumption and NF were simulated as 12 mW and 1.9 dB at 1 GHz respectively. The power supply rejection ratio ranges from 55 dB at 0.15 GHz to 38 dB at 1 GHz due to the unbalanced bandwidth at the two sides of the LNA outputs. The $IIP_3$ and $IIP_2$ performance of the LNA were simulated as 3 dBm and 20 dBm, respectively. The $IIP_2$ performance is limited by the LNA design topology due to its nature of converting single-ended input to differential outputs. A fully differential LNA can exhibit much better $IIP_2$ performance such as the one designed in [33]. However, the fully differential LNA design has the premise that there exists a wideband low-loss balun or differential antenna as a preceding stage. The cost associated with the preceding stage limits the usage of the fully differential LNA.

### 4.3.2 $G_m$ Stage with ESES-based Tunable Gain

The circuit topology of the $G_m$ stage is shown in Figure 4.5, which is similar to the topology in [42]. A fully differential architecture was chosen for better second-order non-linearity performance, which is crucial for direct conversion receiver architecture. A common mode feedback (CMFB) circuit was designed to define the output common mode voltage for the Gm stage. As the Gm stage is DC coupled to the subsequent passive mixer and baseband TIA, this common mode output voltage is hence also the common mode input voltage for the TIA stage.

The amplifier inside the CMFB circuit is a single-stage differential common-source amplifier. This extra amplifier together with the $G_m$ stage forms a two-stage amplifier for the CMFB loop. To stabilize this feedback loop, a Miller compensation capacitor $C_C$ was added.

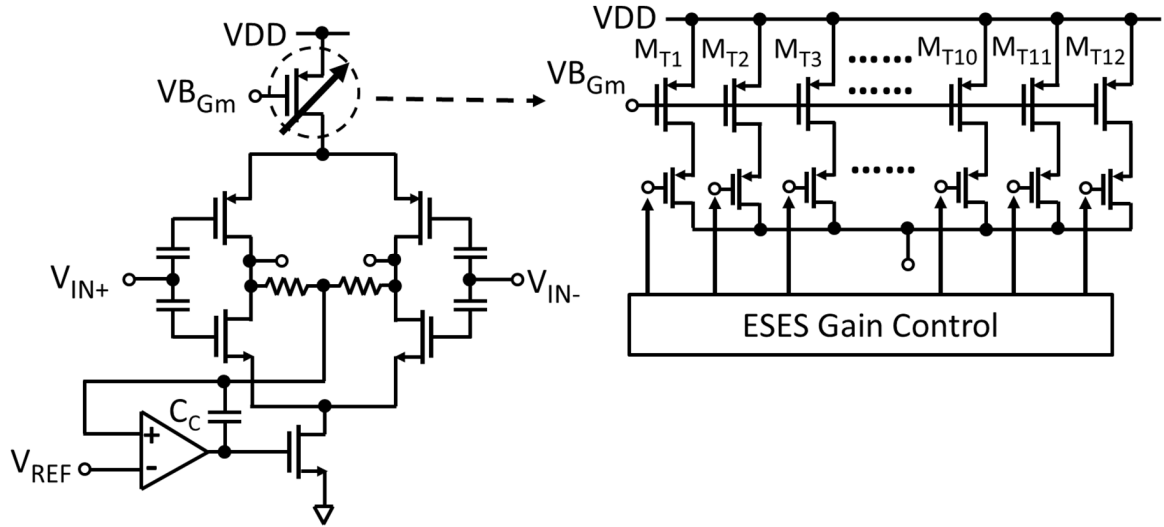

Figure 4.5 $G_m$ stage with tunable gain.

Tunable gain was realized by tuning the size of the PMOS current source tail transistor following the ESES based high-resolution current calibration method (section 3.2). ESES with parameters of N = 12 and K = 6 were applied to sub-transistors $M_{T1-12}$, of which widths were sized with an arithmetic sequence. Switches were placed at the drains of the $M_{T1-12}$ to control the selection of the combination. Each combination has six switches on and six switches off. The total current sets the gain of the $G_m$ stage. The gain tuning range coming from $M_{T1-12}$ must not only cover their own variation, but also cover gain mismatch originating from the PMOS and NMOS input differential pairs, TIAs, and the non-ideal weighting ratio of the resistors in the summing stage. SPICE-level Monte Carlo (MC) simulations show that only 8% of the total gain variation is caused by the PMOS tail current source. The majority of the gain variation is coming from the differential PMOS and NMOS input transistors, as they have the minimal channel length, for bandwidth considerations. With the ESES design method, a sufficiently large tuning range can be created with the PMOS tail current transistor through combinatorial redundancy even though its own randomness is far from dominant. To leave some margin, the gain tuning range was created to cover a six-sigma gain variation seen from the MC circuit simulation. By doing so, applying ESES to a single non-dominant variation source can calibrate gain mismatch from multiple sources. The $G_m$ stage has simulated performance of 8.8 mA/V transconductance gain and 12 dBm $IIP_3$. The CMFB loop has a unity gain bandwidth of 120 MHz and a phase margin of 80°. The power consumption was measured as 3.7 mW with a 1.2 V power supply.

### 4.3.3 Passive Mixer Design

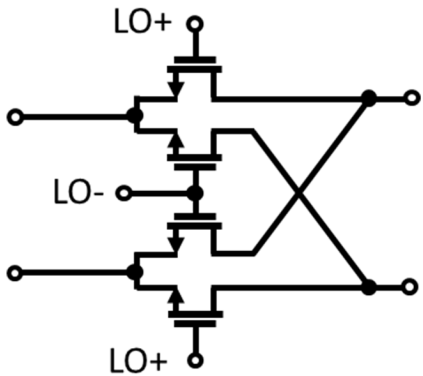

Figure 4.6 Passive mixer circuit design.

The passive mixer design is shown in Figure 4.6. Passive mixers are favored in direct conversion receiver architectures over active mixers because of their zero bias current, which leads to significantly less flicker noise induced by the switching transistors. The input and output is DC coupled to the previous $G_m$ stage and subsequent TIA stage. Differential LO inputs are AC coupled to the LO generator to achieve full rail drive for the switching transistors. Larger switching transistors are preferred due to their small on-resistance. As a result, the previous $G_m$ stage would have less resistive loading and smaller voltage gain resulting in improved linearity.

Moreover, smaller on-resistance means more current is going into the mixer and TIA stages, which can lead to larger gain and better NF performance. However, larger switching transistors also lead to higher power consumption for the driving LO buffers. As a trade-off, the on-resistance was designed to be 50 Ω such that the voltage gain at the $G_m$ output is less than -6 dB for good linearity performance, which also results in a tolerable power consumption for the LO generator and the LO buffer.

### 4.3.4  TIA Design

The TIA design shown in Figure 4.7 was implemented as an OTA with parallel resistor and capacitor feedback for first order baseband filtering. Bypass capacitors were also placed at the inputs of the OTA for both common-mode and differential-mode filtering.

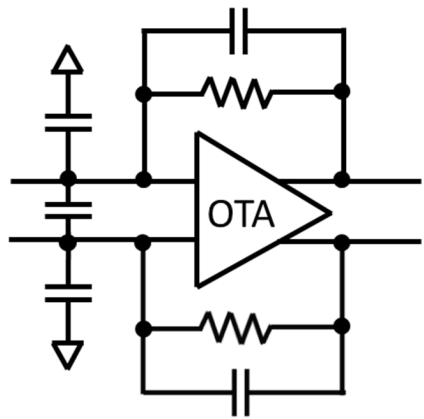

Figure 4.7 TIA design.

The OTA design shown in Figure 4.8 has a typical two-stage OTA topology. The input NMOS transistors were sized with large W/L and with long channel device for noise performance consideration [42]. The first stage has a cascaded stack for only the NMOS devices due to the limitation of the 1.2 V supply voltage. The second stage was designed as a simple common source amplifier stage to achieve a large output voltage swing. The CMFB amplifier was designed as common source amplifier with a diode-connected transistor load. The CMFB loop has three stages which makes stability a challenging issue. To alleviate this issue, the CMFB amplifier was designed to have a very small gain. In addition, the CMFB amplifier was only connected to a fraction of the tail current transistor in the first stage of the OTA. Simulations show that the OTA design has a gain of 63 dB and consumes 3 mW with a 1.2 V power supply. The CMFB loop has a unity gain bandwidth of 115 MHz and a phase margin of 67°.

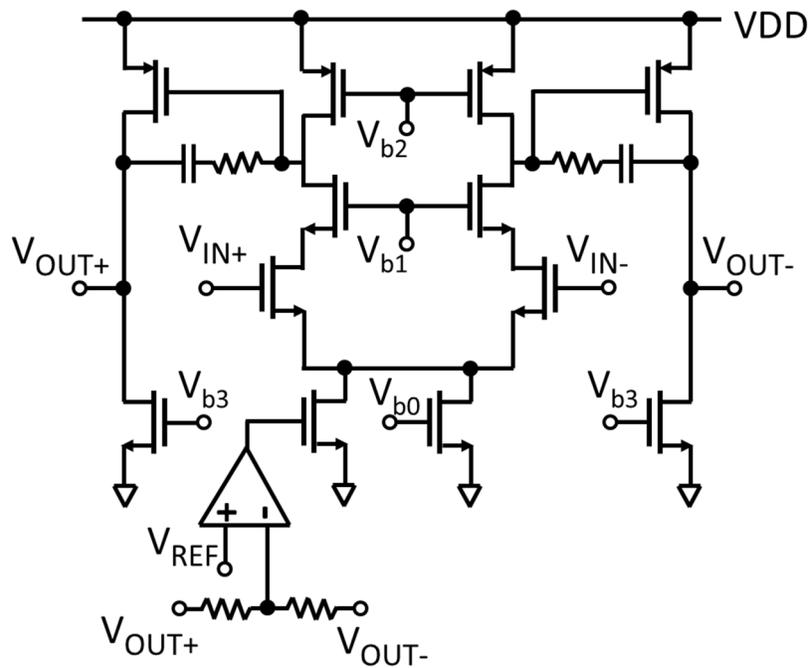

Figure 4.8 OTA circuit design.

### 4.3.5 Multi-phase LO Generator with ESES-based Tunable Phase

The multi-phase LO generator design is shown in Figure 4.9. It consists of a ring of eight differential latches that perform frequency division and generate eight phases. All eight output phases are generated by a single-phase input LO signal. As a result, potential large phase mismatch of the external differential LO signals does not affect the harmonic rejection performance.

To calibrate the 45° phase shift between two adjacent phases, inverters $INV_{1-4}$ were designed to be tunable with the ESES method. This calibration was performed on the clocking signal of the differential latches. Inverters INV$_{5-12}$, which buffer the LO outputs, were also designed to be tunable with the ESES method for differential phase mismatch calibration for even-order HR.

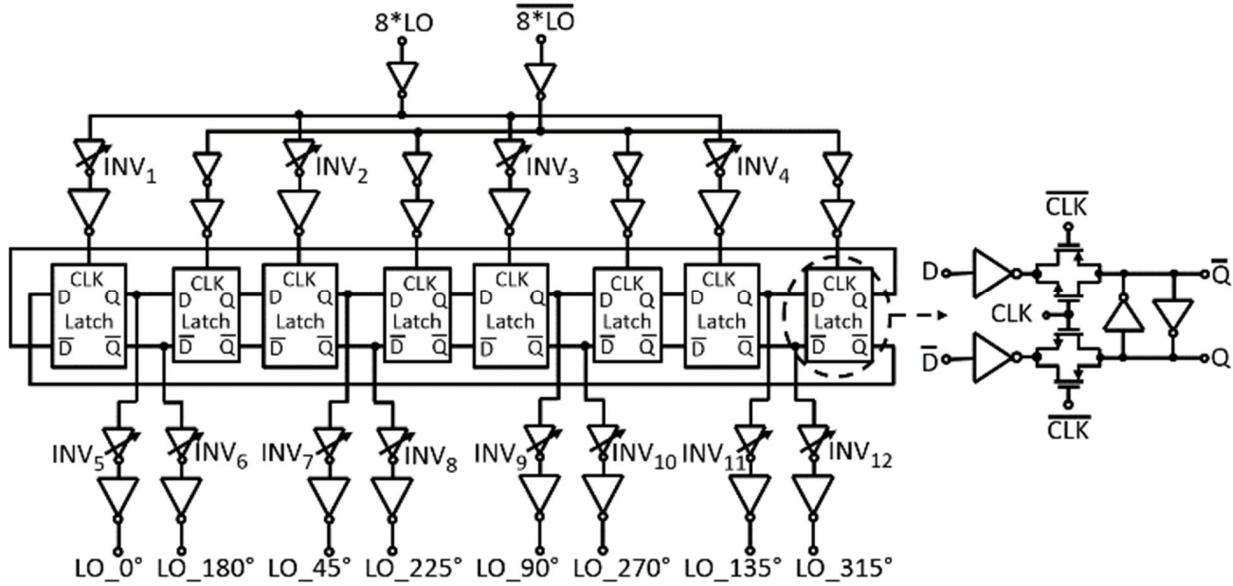

Figure 4.9 Divide-by-8 multi-phase LO generator with tunable phase.

The tunable inverters were implemented as shown in Figure 4.10. The ESES ESES-based high resolution phase delay calibration method (section 3.3) was applied by breaking the PMOS and NMOS transistors into a parallel combination of smaller transistors (N = 12 and K = 6 adopted) with sizes of an arithmetic sequence. By selecting different combination of the sub-transistors, different rising or falling edge delay can be achieved as the driving force is determined by the overall transistor width of the sub-transistors that are selected. Two versions of tunable inverters were used in this design. For HRR$_2$ calibration of differential phases, where both the rising and falling edges are important, INV$_{5-12}$ both the PMOS and NMOS transistors can be tuned. For HRR$_3$ calibration of 45° phase mismatch, only the rising edge is critical, thereby limiting the need for tuning to the NMOS transistors in INV$_{1-4}$. To successfully calibrate the phase delay mismatches, the tuning range provided by a single inverter (INV$_{1-12}$) must cover the phase delay variation coming from a logic chain. In simulation, 25% of the total clock delay variation is coming from $INV_{1-4}$ and 45% of the total differential phase mismatch is coming from $INV_{5-12}$. By using ESES, we were able to design the sub-transistors with a wider spread of transistor widths (i.e., larger common difference for the arithmetic sequence for sizing) to cover a wider calibration range. In this design, with proper element sizing, sufficient tuning ranges were created to cover a worst case six-sigma phase variation.

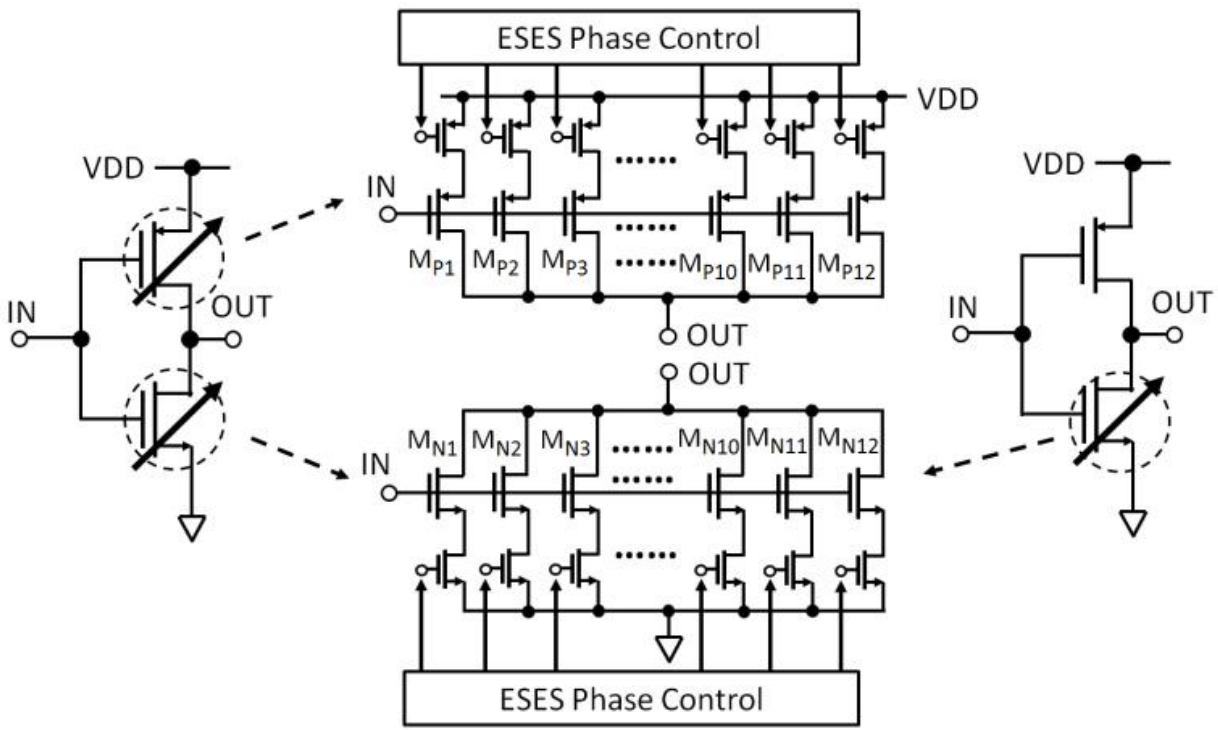

Figure 4.10 Inverter design with tunable delay.

## 4.4 ESES Calibration Process

### 4.4.1 ESES Calibration for HR Receiver Overview

All calibration steps start with injecting a specific harmonic test tone at the receiver input that emulates harmonic interference and then measuring its power at the receiver output with an ADC. The output power of one ESES-enabled design choice is then converted to an HRR value. Different ESES design choices are tested and the design choice that has the best HRR performance is selected as the final configuration. In this proof-of-concept implementation, the harmonic test tone is supplied by external RF signal generator, and the ADC and digital signal processing (DSP) circuits are realized with off-chip data acquisition hardware and a PC respectively. For a complete on-chip implementation, a baseband ADC would be present for a complete receiver design such that the harmonic tone injection could be realized by the on-chip transmitter. The only circuit overhead for ESES-based calibration for a transceiver design is the extra DSP units and memory cells for storing the final ESES configurations. This digital circuit overhead associated with ESES calibration will scale with CMOS technology.

### 4.4.2 Even-order HRR Calibration Flow

As all even-order HR has the same condition for ideal HRR value, which is perfect 180° phase shift within a pair of differential LO phases, the even-order HRR calibration can be performed on $HRR_2$ only and the resulting optimal ESES design choice for $HRR_2$ can also provide optimal $HRR_{4,6}$ values.

Even-order HRR calibration process started with injecting a second order harmonic tone. Only one branch among the four branches in the receiver was turned on at a given time and a pair of tunable inverters associated with this branch (e.g. $INV_{5,6}$ for the branch of 0°/180° branch) underwent ESES tuning. The corresponding $HRR_2$ value for each tested ESES design choice was measured and the best one was selected as the final result. The calibration was then repeated for all four branches. This results in not only the optimization of $HRR_2$ for the entire receiver including both I/Q path, but also all other even order HRR, as previously discussed.

### 4.4.3  3$^{rd}$ and 5$^{th}$ Order HRR Calibration Flow

Calibration of the 3$^{rd}$ and 5$^{th}$ order HHR was performed by injecting a 3$^{rd}$ order harmonic tone. The calibration method is to calibrate gain errors at a low frequency and calibrate, phase errors at the working frequency. This method is based on first order approximations that gain errors are not a function of frequency and phase errors scale down at lower frequency. Assume that the frequency band under calibration is $F_0$. ESES gain calibration was first performed at the lowest frequency band for $G_m$ stages, where the phase mismatch was smallest. After achieving an optimized $HRR_3$ value with gain calibration only in the lowest frequency band, the values were stored. Phase calibration in band $F_0$ was then performed using these stored gain calibration results. The phase calibration applied ESES to $INV_{1-4}$. Assuming the lowest frequency band is free of phase mismatch, then gain and phase mismatch have been calibrated independently resulting in optimization of both $HRR_3$ and $HRR_5$ simultaneously. To validate this assumption, an iterative process is needed for gain calibration in the lowest frequency band and phase calibration at $F_0$. From chip measurements, two iterations have proven to be sufficient to make $HRR_{3,5}$ converge to their optimal values. The calibration of I and Q paths were performed sequentially. Although the general flow is the same for both I and Q paths, the I path calibration involves $G_m$ stages and $INV_{1-3}$ in 0°/180°, 45°/225° and 90°/270° branches, while the Q path calibration only deals with the remaining $G_m$ stage and $INV_4$ in 135°/315° branch. The aforementioned calibration flow would optimize the 3$^{rd}$ and 5$^{th}$ order HRR simultaneously for the receiver. Furthermore, the previously performed even-order HRR calibration results still hold.

This is because in 3$^{rd}$ and 5$^{th}$ order HRR calibration, only the clock input delay for the differential latch is changed, which has an equal impact on the latch's differential outputs. For example, 0° and 180° phases are delayed equally during 3$^{rd}$ and 5$^{th}$ order HRR calibration, but they still maintain 180° phase shift between each other after even-order HRR calibration is done.

## 4.5 Harmonic Rejection Receiver Testing

### 4.5.1 Measurement Setup

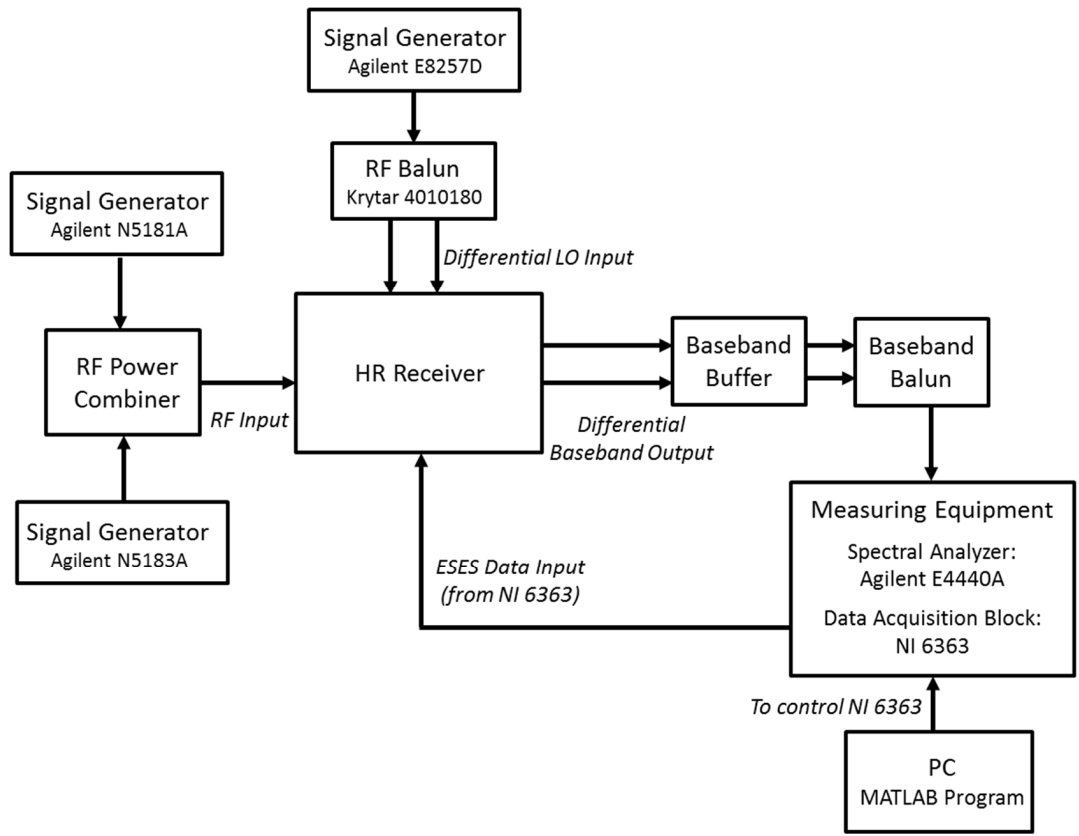

Figure 4.11 Measurement setup diagram.

The HR receiver chip was assembled in a QFN-64 package and mounted on a regular FR-4 printed circuit board (PCB). A measurement setup diagram is shown in Figure 4.11. The differential LO input was supplied by an Agilent signal generator followed by a wideband RF balun. For the one tone test and the harmonic rejection test, only one signal generator was used at the RF input. The linearity test requires two tones and thus two signal generators and a following RF power combiner were used at RF input. For the HR receiver baseband outputs, unity gain buffers were designed on a PCB to drive the large capacitive load or 50 Ω resistive load of the test equipment. For noise measurement and linearity measurements, an Agilent spectrum analyzer with noise figure measurement function was used. For harmonic rejection measurements and calibration, a National Instruments data acquisition block was used to digitize the output signal. The signal was analyzed externally to calculate the harmonic rejection ratio for a specific ESES design choice. Different design choices for ESES calibration were also supplied by the data acquisition block to the HR receiver.

### 4.5.2 Measurement Results

The wideband RF receiver with ESES-based HR calibration was designed in 65 nm bulk CMOS technology from Global Foundries. The entire die area is 2 mm by 2 mm, while the core circuit area is 0.72 mm$^2$ (die photo shown in Figure 4.12). The divide-by-eight multi-phase LO generator limits the RF range of this design to 1 GHz. The total gain is 48 dB, with less than 1 dB variation within 0.15 GHz to 1 GHz as the RF bandwidth of the receiver was designed to be 3 GHz which is much higher than 1 GHz. The analog power consumption with a 1.2 V supply is 45 mW, which consists of 12 mW for LNA, 15 mW for the $G_m$ stages, and 18 mW for the TIA stages. The LO generator consumes 19 mW at 1 GHz and the power of this block scales down almost linearly as the frequency goes down. The noise figure was measured as 2.8 dB at 1 GHz which gradually increases to 3.2 dB at 0.15 GHz due to flicker noise at lower frequency. The out-of-band IIP$_3$ (OB-IIP$_3$) was measured at -7 dBm, which is limited by the large gain of the LNA as well as IIP$_3$ performance of the $G_m$ stage. Out-of-band IIP$_2$ (OB-IIP$_2$) was measured at +22 dBm, which is limited by the IIP$_2$ performance of the balun-LNA. The baseband 3dB bandwidth is 13 MHz.

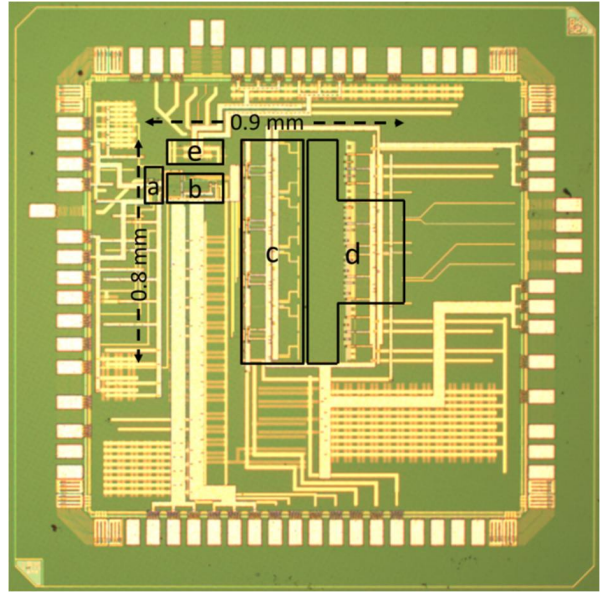

Figure 4.12 Harmonic rejection receiver die photo (a. LNA; b. Gm stages; c. TIA branches; d. harmonic recombination for I/Q path; e. LO generator).

Harmonic interference up to 3 GHz was injected for HRR testing during the chip measurement. The HR performance of one typical chip is shown in Figure 4.13. For even-order HRR, only $HRR_2$ is shown, as $HRR_{4,6}$ have similar if not better performance than $HRR_2$. The results in Figure 4.13 show the HR before and after calibration at each frequency point on the plot. More than 80 dB of $HRR_{2,3,4,6}$ and more than 70 dB of $HRR_5$ were achieved after calibration. The calibration performed at a single frequency was further evaluated over a wide frequency range. Figure 4.14 shows that a single calibration at 0.75 GHz resulted in more than 70 dB HRR from 0.15 GHz to 0.75 GHz. Comparing performances shown in Figure 4.13 and Figure 4.14, the even-order harmonic rejection ratios do not have any degradation while only performing calibration at one single frequency. This is expected as the differential LO phase errors are caused by the inverter delay errors which are not frequency dependent for the first order approximation. For $HRR_3$ and $HRR_5$, we can see that $HRR_3$ has noticeable degradation while $HRR_5$ has almost no degradation when calibration is only carried out in one frequency.

This might be because after gain error calibration, there still exist a small portion of gain errors. And the remaining gain errors are then cancelled out by the phase errors during the phase error calibration. As the test tone is a third order harmonic, this cancellation of the remaining gain errors can only benefit $HRR_3$ at the frequency that the calibration is performed.

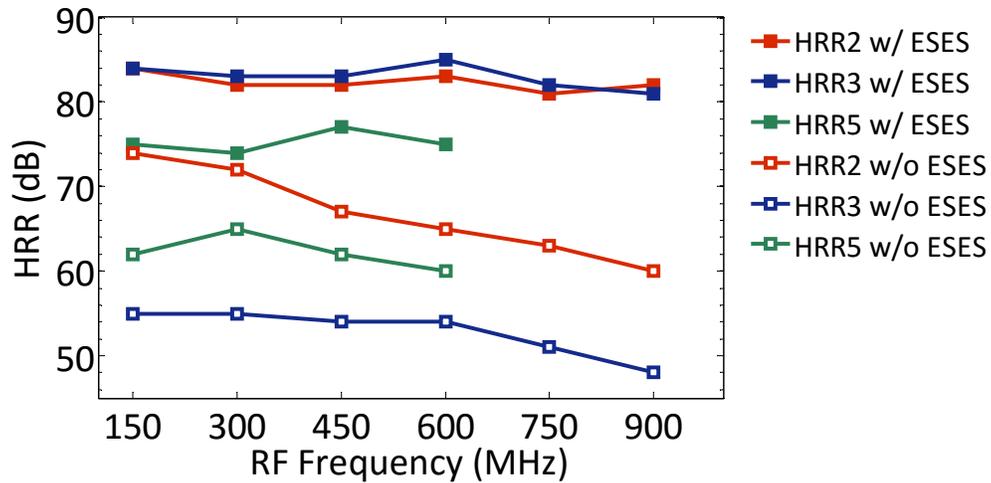

Figure 4.13 HRR performance w/o and w/ ESES calibration.

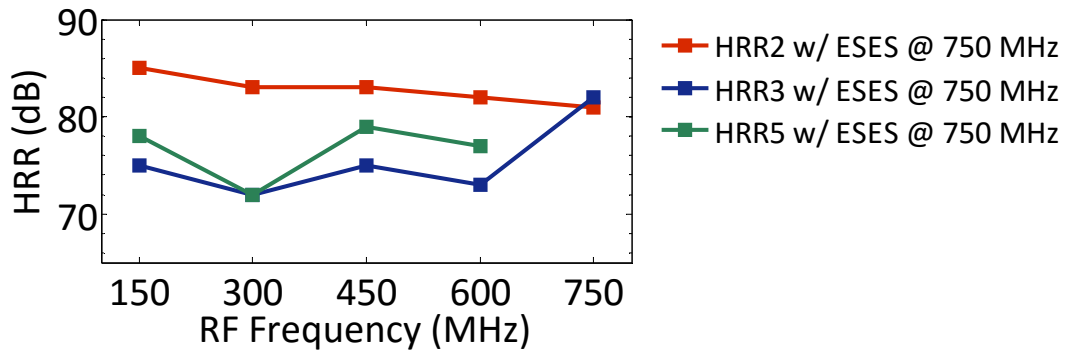

Figure 4.14 HRR Performance after ESES-based calibration at 750 MHz

Table 4-1 shows the comparison with the prior works of HR receiver design with a high HRR. In [39], the HR scheme with adaptive interference cancellation achieved 82.5 dB HRR3 and 81 dB HRR5 but not simultaneously. Since the HR was mostly done in digital domain with two extra baseband ADCs, this work is not included in Table 4-1 for comparison.

Table 4-1 Comparison with prior works.

| | This work | [33] ISSCC '09 | [35] ISSCC '14 | [37] VLSI '13 | [38] JSSC '13 |
|---|---|---|---|---|---|
| RF Range (GHz) | 0.15 to 1 | 0.4 to 0.9 | 0.048 to 0.87 | 0.4 to 6 | 0.04 to 1.002 |
| Gain (dB) | 48 | 34 | 36 (max) | 70 (max) | - |
| NF (dB) | 3.2 | 4 | 3.1 | 1.8 - 2.4 | 3 |
| OB-IIP$_3$ (dBm) | -7 | +18 | - | +3 | - |
| OB-IIP$_2$ (dBm) | + 22 | + 56 [1] | + 59 [1] | + 80 [2] | + 20 |
| HRR$_3$ (dB) | **> 80** | > 60 | > 60 [3] | > 70 | > 65 |
| HRR$_5$ (dB) | **> 70** | > 64 | > 60 [3] | > 55 | > 65 |
| HRR$_{2,4,6}$ (dB) | **> 80** | > 62 | - | - | > 65 |
| Power (mW) | 64 | 60 | 183 | 40 [4] | 440 [5] |
| Area (mm$^2$) | 0.72 | 2.7 | 1 | 0.6 [4] | 5.6 [5] |
| Technology | 65 nm | 65 nm | 130 nm | 28 nm | 80 nm |

[1] differential RF input, [2] differential RF input with IIP$_2$ calibration,
[3] exclude RF filtering, [4] include ADC, [5] include PLL + ADC + DSP

The second-order linearity performance of this design is limited by the single-ended RF input for the receiver. The third-order linearity performance of this design suffers from the large LNA gain and succeeding G$_m$ stages. Improved linearity can be realized in wideband receiver architectures by utilizing differential RF input for second-order linearity performance while at the cost of preceding differential stage, and through the use of a low noise transconductance amplifier as the first stage [33] or with a passive mixer first stage [43], for the third-order linearity performance while at the cost of increased NF. The ESES-based HRR calibration technique implemented in this prototype design can be applied to these wideband receiver architectures to achieve both best-in-class HRR and ultra-high linearity, thereby producing a wideband receiver with increased tolerance to out-of-band blockers.

## 4.6 Summary

The HR receiver design shown in this chapter demonstrates the utility of ESES design method in a complex RF design. As ESES-based gain and phase calibration were applied to the HR receiver at several locations, best-in-class HRR values have been obtained after calibration. The ESES-based design method has been proven to provide high calibration resolution for achieving high matching accuracy while creating a sufficiently large calibration range to accommodate numerous random variation sources in circuit design.

# 5  Current-Steering D/A Converter Design

In this chapter, as a second example of our proposed calibration methods, we present a current-steering D/A converter (CS-DAC) design with ESES-based calibration for amplitude error and timing error to improve CS-DAC linearity performance.

## 5.1  Current-Steering DAC Overview

### 5.1.1  Current-Steering DAC Linearity Enhancement Methods

Current-steering DACs (CS-DACs) are the most suitable DAC architecture when both high speed and high resolution are needed, such as for video and radio transmitter applications. The linearity of CS-DACs is limited by amplitude errors of current sources at low frequency and timing errors at high frequency.

Amplitude errors of CS-DACs lead to static linearity problem, and can be improved by upsizing current source transistors at the cost of significant circuit area increase for high resolution applications [8]. Many methods [19] [20] [22] [44] [45] have been proposed for calibrating amplitude errors, thus achieving better static linearity. The basic concept of amplitude error calibration for segmented CS-DACs was summarized as a two-phase process in [44]; namely, sensing errors and adjusting errors. The sensing phase can be done with just one current comparator [20] [22]. In [20], a calibration method using floating current sources for adjustment was proposed at the cost of complex analog circuits that are more challenging to implement as process technology scales and supply voltage decrease. In [22], a digital adjustment method of attaching calibration DACs (CALDACs) for every calibrated current source was proposed that results in a large circuit area for the CALDACs. Alternatively, an ADC can be used for sensing, but that generally requires significant calibration circuitry overhead, such as an ADC-DSP-CALDACs loop [19] [45]. A non-traditional calibration method was proposed in [44] for segmented CS-DACs. The static linearity was improved by adjusting the switching sequence of the current cells to minimize the accumulation of current errors. Minimal overhead was incurred in terms of one additional current comparator. However, the static linearity improvement was relatively small.

In recent years some work [46] [47] has focused on optimizing both amplitude errors and timing errors (consists of delay errors and duty cycle errors) such that much better dynamic performance at high frequency can be achieved. Calibration techniques based on optimizing accumulated errors through sorting were used. Instead of optimizing one-dimensional amplitude errors as in [44], these designs optimized three-dimensional errors, namely amplitude errors, delay errors and duty cycle errors together. However, during optimization a weight between amplitude errors and timing errors has to be determined. This results in a trade-off between low frequency and high frequency linearity as shown in [46]. Also, since this is a hyper-dimensional optimization problem, the optimized overall error is a rather compromised result compared to the case where less error sources are present.

To achieve further improved linearity performance over the entire Nyquist band, amplitude errors and timing errors need to be optimized independently. Moreover, analog circuit overhead should be kept as minimum such that the calibration method can be friendly for process technology node scaling. For amplitude error, most of the existing calibration methods, such as the aforementioned [19] [20] [22] [45], involve high analog circuit overhead in the form of large CALDACs, which occupy significant chip area that only scales down with the improvements of matching property. And the method with least analog circuit overhead as shown in [44] has limited improvement as compared to other methods. For timing error calibration, only two works [46] [47] have addressed this issue till now. They shared the same calibration method which is through optimizing the switching sequence of the thermometer coded MSB cells. Similar to the amplitude error optimization method in [44], the error from each individual thermometer coded cell does not change after optimization. We can expect larger improvements of the timing error if the timing error of each thermometer coded MSB cell can be individually calibrated.

### 5.1.2 Previous Work on SES-based Current-Steering DAC

An SES-based current-steering DAC was designed previously to address the amplitude errors for improving the static linearity [11]. As the static linearity of segmented CS-DACs is primarily attributable to current matching of the unary current cells (UCCs) in several-bit thermometer coded MSBs (most significant bits), SES design method can be applied to these UCCs by forming each cell with N sub-cells, but only activating K sub-cells for operation. With this combinatorial redundancy, the current cells can be digitally configured to have very small errors relative to the nominal value. If every UCC for the MSBs can be selected as a combination with a much reduced error, the current matching of the CS-DAC is significantly improved. As the UCC under SES calibration is the dominant variation source for the overall current matching, SES is applicable in this application even though the calibration range is limited.

Since the SES method is only applied to the current cells without changing normal DAC operation, the CS-DAC after self-healing acts like the design that has intrinsic high matching properties. Therefore, this SES-based self-healing method provides a solution for CS-DAC designs to independently optimize amplitude errors, hence static linearity.

A self-healing control loop can be built to search and select a combination for each UCC with a much reduced amplitude error. To do so requires an on-chip target for comparison. Consider the summation of all binary weighted LSBs' (least significant bits) current cells plus one more LSB current value specified as $I_{ref}$. This value $I_{ref}$ has the same nominal value as the UCCs, therefore it represents an ideal target. One tiny current source, $I_{tiny}$, is further introduced to create a very small target window for UCCs, namely [$I_{ref}$, $I_{ref}+I_{tiny}$]. A digital controller is constructed to search for one combination of K sub-cell currents that falls into that window. The cell level SES-based design is constructed such that this combination can be found via random search with a trials limit imposed to bound searching time. Although this combination might not be the one having least error, searching for this local optimal solution simplifies hardware implementation to just one comparator rather than a high resolution ADC. It is noted that the value of $I_{tiny}$ is a trade-off between the "optimality" of the UCC values and the rate of successful SES search. This was mathematically modeled and evaluated as part of the design process.

The challenge is to make all of the UCCs for the MSBs successfully fall into the same target window. From simulation we noted that the cell-level SES success rate drops quickly as the target window deviates from the nominal value of UCCs. Therefore, to improve the self-healing performance, the nominal value of UCCs is designed with tunability. This is achieved by applying a top level SES on the UCCs' biasing current mirror that is also split into a set of selectable transistors just like UCCs. Through a trial process, the top level SES operation finds out one combination for the biasing current that every UCC can successfully find one combination falling in the small target window. A limit is also placed on the number of random search trials for top level SES.

Additional backup UCCs corresponding to traditional redundancy are added to boost the successful rate of self-healing. The overall SES-based self-healing logic flow is shown in Figure 5.1.

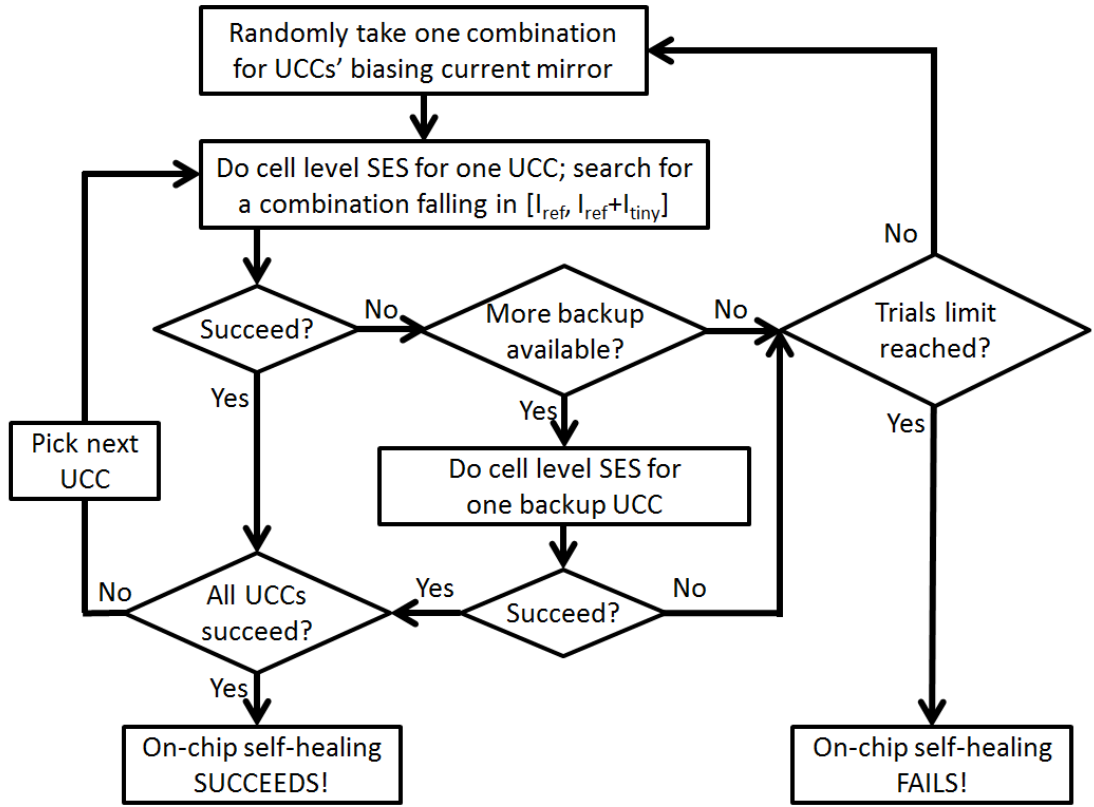

Figure 5.1 Overall SES-based self-healing logic flow.

This SES-based self-healing flow can reduce the UCC current deviations for thermometer coded several-bit MSBs, while the remaining LSBs current errors are relatively unchanged. Therefore, after self-healing operations for MSBs, the amplitude errors would be dominated by uncalibrated LSBs. To maximize the overall static linearity performance, the errors coming from LSBs are optimized by sizing the channel lengths of the LSBs several times larger than those for the MSBs. By doing so, LSBs can have improved intrinsic matching properties, thereby not dominating overall amplitude errors after self-healing on MSBs. Area overhead of oversizing LSBs is far more affordable than oversizing MSBs, since LSBs only occupy a small portion of overall area.

Figure 5.2 shows the implementation of the aforementioned on-chip self-healing flow. The primary analog circuit overhead is one current comparator and the digital circuit overhead is an on-chip self-healing controller.

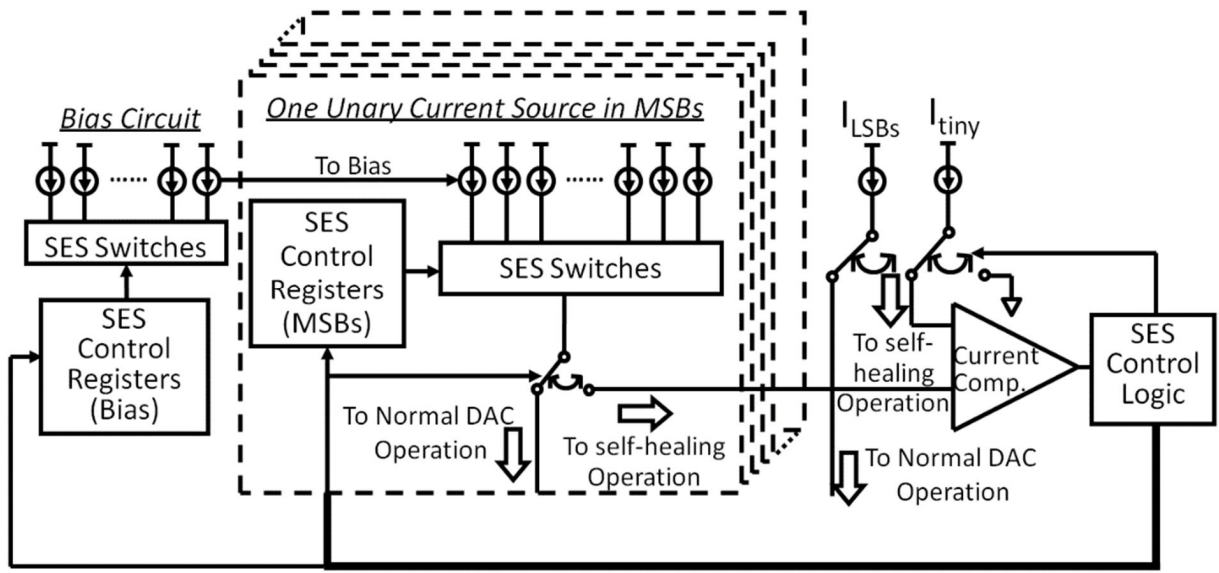

Figure 5.2 System diagram of SES-based self-healing CS-DAC.

The UCC design needs to be modified into configurable circuit components for selection. A typical UCC design is shown in Figure 5.3 (a), and our SES-based UCC design is shown in Figure 5.3 (b). This modified design has split current source transistors, shown as $M_{0\_1}$ to $M_{0\_16}$ in Figure 5.3 (b), that are controlled by a set of switches, $S_1$ to $S_{16}$. Each split current source has a transistor size of 4 μm/1 μm and a nominal current value of 19.53 μA. During SES-based searching, 8 of them are turned on by the control registers. The switches are operated in the deep triode region, which consumes very little voltage headroom. A cascode transistor $M_1$ is placed at the drains of the current source transistors for shielding the capacitance of that node. This also helps to significantly lower down the impact of having different parasitic capacitance contributions from MSBs and LSBs by using different channel lengths for current sources. One more current path for comparison purpose was created that is also controlled by switches operated in the deep triode region. The switching pair design is the same as in a typical design.

The standard deviation for each of the 6-bit MSBs' UCCs was simulated as 0.53 μA in this design.

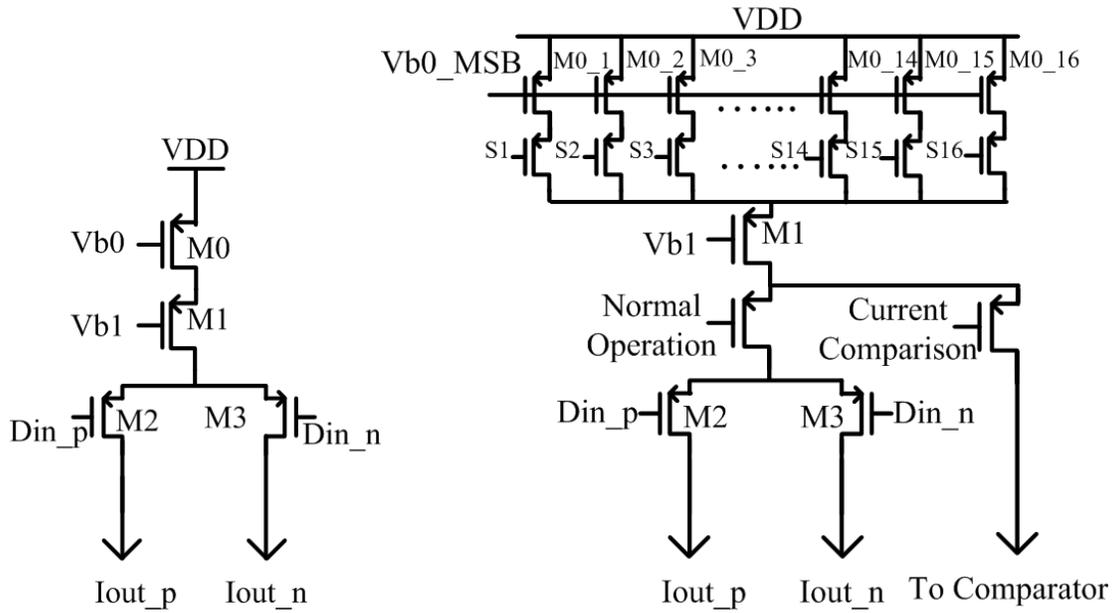

Figure 5.3 (a) Typical UCC Design; (b) SES-based UCC Design.

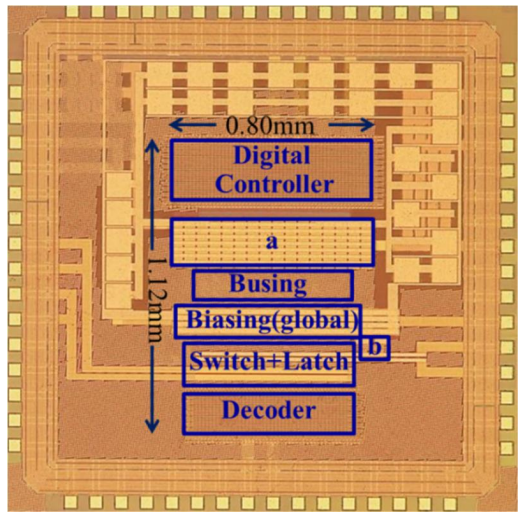

Figure 5.4 Die photo of self-healing 14-bit CS-DAC. a. current source array + local biasing + dummies. b. current comparator.

A prototype 14-bit CS-DAC design with 6-8 segmentation was implemented in 130 nm CMOS technology to demonstrate the proposed SES-based self-healing method. The die photo is shown in Figure 5.4. The core area was measured as 0.9 mm$^2$, with less than 0.1 mm$^2$ occupied by the current source array. The self-healing overhead cost was measured as a current comparator in 0.01 mm$^2$ and a digital controller in 0.16 mm$^2$.

A total of 5 chips were tested and all of the chips succeeded in SES-based self-healing operation. This meets our expectation as the success rate was simulated to be more than 99.7% for this implementation. The static linearity improvement of one typical chip is shown in Figure 5.5. Before self-healing, $INL_{max}$ was measured as 6.28 LSB and $DNL_{max}$ was measured as 2.32 LSB. After self-healing, $INL_{max}$ was reduced to 0.64 LSB, and $DNL_{max}$ was reduced to 0.66 LSB.

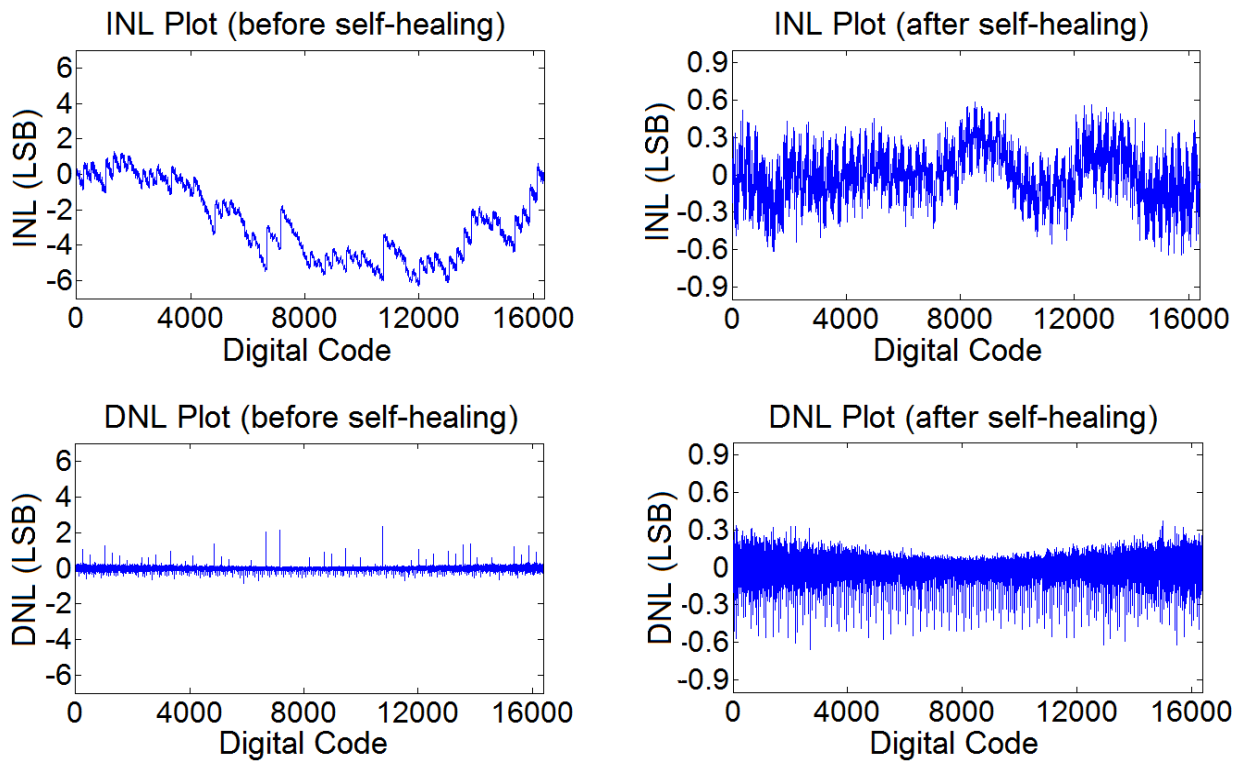

Figure 5.5 INL, DNL plots before SES-based self-healing and after SES-based self-healing.

Table 5-1 Performance summary of the SES-based CS-DAC.

| Resolution | 14 bits |
|---|---|
| Max Sampling Rate | 200 MS/s |
| Full-scale Current | 10 mA |
| Output Swing | 1 V (differential) |
| Supply Voltage | 1.5 V |
| $INL_{max}$ | 6.28 LSB (before)  0.64 LSB (after) |
| $DNL_{max}$ | 2.32 LSB (before)  0.66 LSB (after) |
| SFDR at 200 MS/s | 73 dB (before)  85 dB (after)  at $f_{sig}$ = 2 MHz<br>64 dB (before)  67 dB (after)  at $f_{sig}$ = 10 MHz<br>48 dB (before)  48 dB (after)  at $f_{sig}$ = 50 MHz |
| Power Consumption | 18 mW (analog); <10 mW (digital) |
| Core Area | 0.9 mm$^2$ in 130 nm CMOS technology |

A performance summary of the SES-based CS-DAC design is shown in Table 5-1. This previous CS-DAC design with SES-based amplitude error calibration succeeded in achieving high static linearity after calibration. However, this implementation has the following drawbacks:

1) Due to the low calibration success rate of the SES design method, large N and K values (N = 16, K = 8) were selected. This increases the number of design choices significantly while at the cost of large number of calibration operations.

2) The transistor spatial errors are independent of transistor random mismatch. Due to the small calibration range of the SES design method, the spatial errors cannot be covered by SES-based tuning range that is generated by random mismatch only. To counter this, SES-based current cell has to be broken into small pieces and distributed across the entire current cell array to make the spatial error also contribute to the overall SES-based tuning range. But this results in a complicated routing scheme for the current cell array and thereby extra interconnect parasitic for the current cells which degrades dynamic performance of the DAC.

3) The timing errors of the CS-DAC were not calibrated. Using SES-based method alone is not suitable for calibrating timing errors as the timing errors usually are not determined by a single variation source. Without timing error calibration, the linearity of the CS-DAC and the dynamic performance drops significantly at high frequency input.

A new CS-DAC design with ESES-based calibration is presented in the next section that can address the drawbacks of the SES-based CS-DAC design.

## 5.2 ESES-based Current-Steering DAC Design

In order to address the issues of the SES-based CS-DAC design as discussed in section 5.1.2, we propose an ESES-based CS-DAC design which uses ESES design method to calibrate both amplitude errors and timing errors of the CS-DAC, such that we can potentially achieve high linearity across the entire Nyquist band.

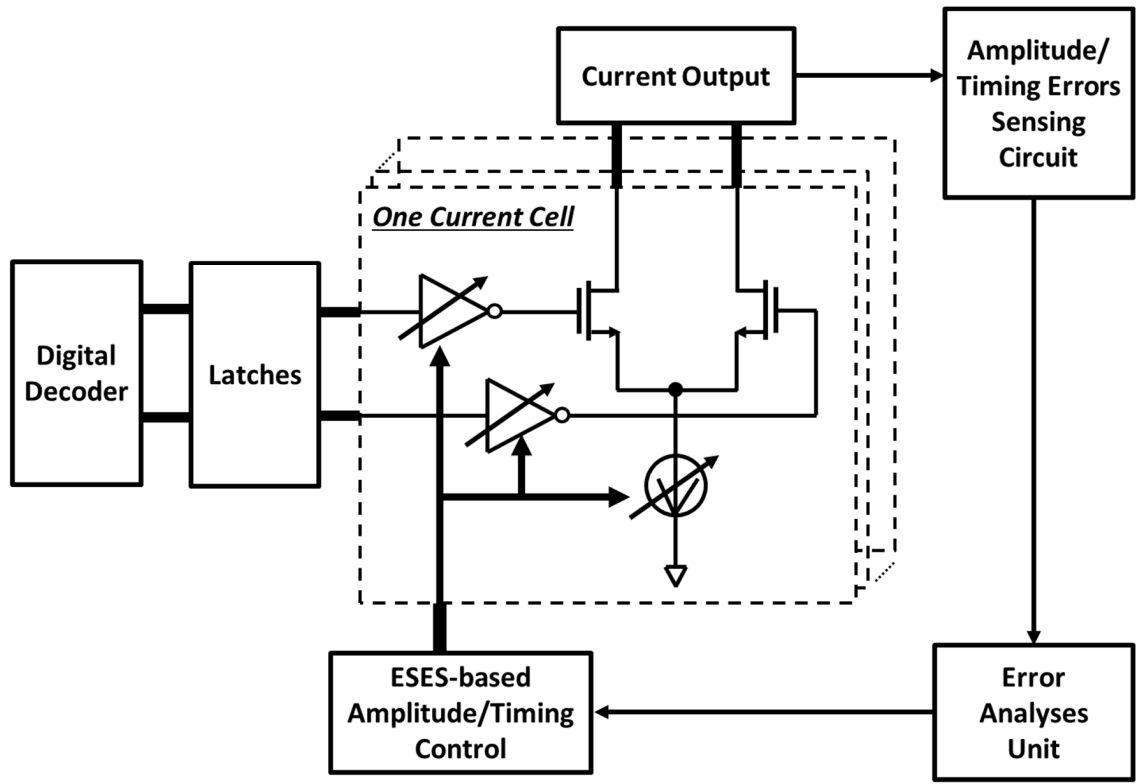

Figure 5.6 Conceptual ESES-based CS-DAC design

A conceptual drawing of the ESES-based CS-DAC design is shown in Figure 5.6. This shows the circuit diagram for one current cell in the thermometer coded MSBs. The current source of this cell is designed to be tunable by ESES-based amplitude control. The switching timing of this cell is also designed to be tunable by ESES-based timing control. The current output of this cell has a path that goes to error sensing circuit. The sensed error is then processed by an error analyses unit, which mainly consists of analog to digital converter (ADC) and digital signal processing unit. For our prototype demonstration, the error is analyzed using an off-chip data acquisition block and a PC, which is similar to the implementation as shown in section 4.5.1. The off-chip error analyses unit then feeds back to the ESES-based amplitude/timing control units for on-chip control.

The amplitude/timing sensing scheme was proposed in [46]. To sense the amplitude/timing errors between two current cells, the first step is to switch the two current cells together such that the current outputs have a measurement output frequency $f_{meas}$. The next step is to take the current output difference of the two cells. Figure 5.7 shows the sensing scheme for amplitude error. If there only exists amplitude error between current cells, the difference of the two cells' current outputs has a fundamental frequency of $f_{meas}$ and this amplitude error signal has the same phase as the two current cells' outputs. Then we multiply this amplitude error signal with a modulation signal by using a mixer. If the modulation signal also has the fundamental frequency of $f_{meas}$ as well as the same phase as the two current cells' outputs, just as shown in Figure 5.7, then the mixer's output has a DC value which is proportional to the magnitude of the amplitude error. Therefore, through the DC value of the mixer's output, we can sense the amplitude error between two cells.

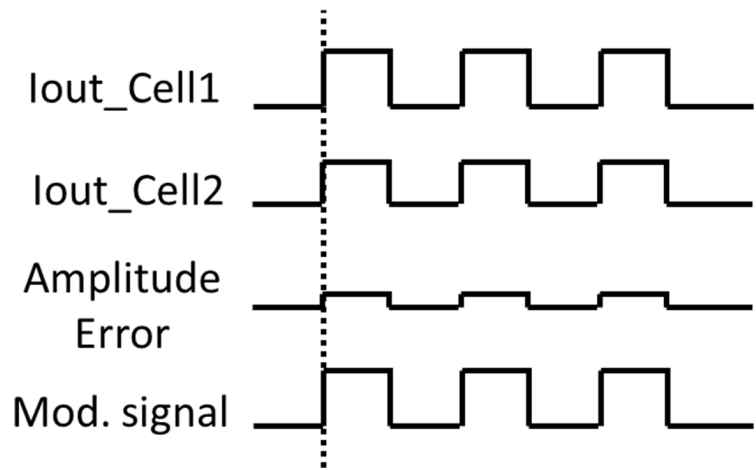

Figure 5.7 Amplitude error sensing scheme.

Similarly, if there only exists delay error between two current cells, then the difference of the two cells' current outputs, which is the delay error signal, has a fundamental frequency of $f_{meas}$ and a phase shift of 90° from that of the current cells as shown in Figure 5.8. The phase of the modulation signal should now also be 90° off from that of the two current cells' outputs, such that the DC output of the mixer can quantify the delay error between two current cells. It is noted that during the process of amplitude error sensing, the delay error ideally would not have any DC response at the mixer output because the delay error is orthogonal to the modulation signal for the amplitude error sensing process. This is also true for the impact of the amplitude error during the process of delay error sensing. This means amplitude error and delay error can be sensed separately by using this sensing scheme.

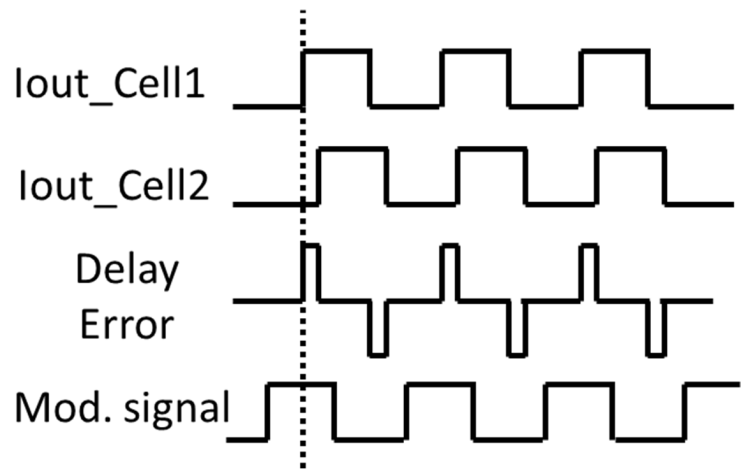

Figure 5.8 Delay error sensing scheme.

For the duty cycle error between two current cells, the error signal has a frequency of $2*f_{meas}$ as shown in Figure 5.9. Therefore, the modulation frequency should also has a fundamental frequency of $2*f_{meas}$ in order to sense the duty cycle error through the DC output of mixer. It is also noted that during the process of amplitude/delay error sensing, the duty cycle error ideally does not generate DC output because the modulation signal of amplitude/delay error sensing process does not have even-order harmonic components. This is also the case for the impact of amplitude/delay error during the process of duty cycle error sensing. Therefore, duty cycle error can be sensed separately from amplitude/delay error.

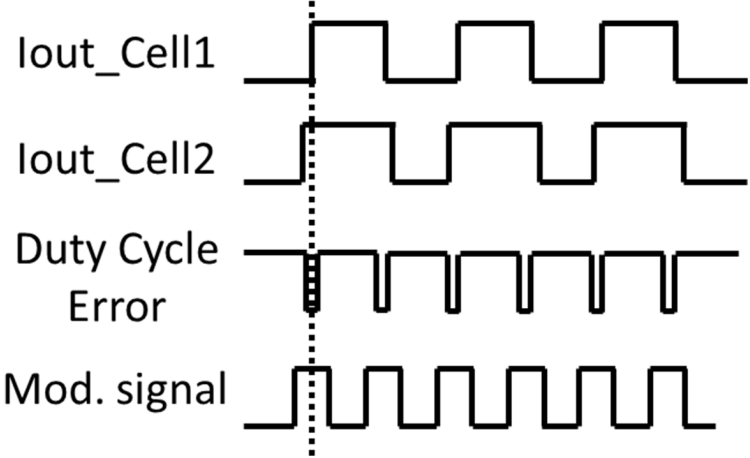

Figure 5.9 Duty cycle error sensing scheme.

## 5.3 Design Implementation

A 14-bit CS-DAC with 6-8 segmentation (6-bit thermometer coded and 8-bit binary weighted) was implemented in the Samsung 28nm bulk CMOS process. ESES-based current and delay calibration methods were employed in this design by adapting the unary current cell design for the 6-bit thermometer coded MSBs and latch design for each unary current cell respectively. The details of these two critical designs are presented in this section. The implementation of the amplitude/timing error sensing circuit is also included in this section.

### 5.3.1 Unary Current Cell Design with ESES-based Tunable Current

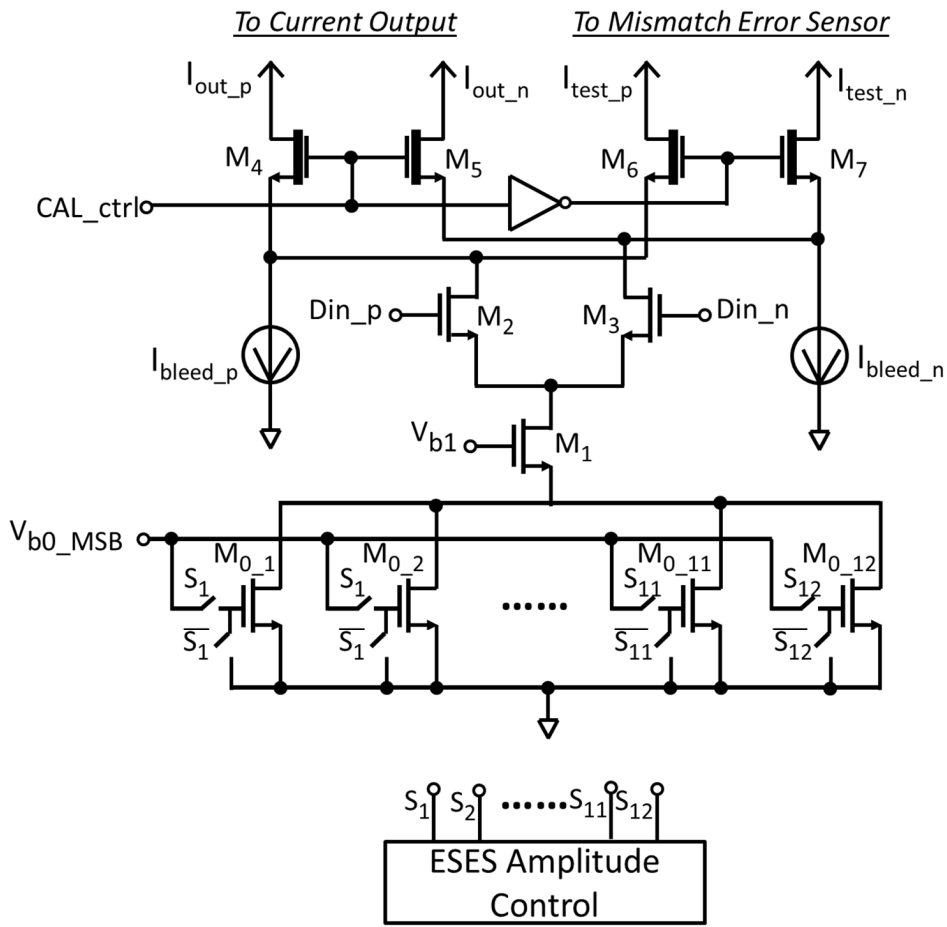

Figure 5.10 ESES-based unary current cell Design.

Figure 5.10 shows the implementation of the ESES-based unary current cell (UCC) design. NMOS transistors were chosen in this implementation for high speed and low power considerations. The current source of the UCC is split into an array of 12 sub-transistors $M_{0\_1}$, $M_{0\_2}$, ..., $M_{0\_11}$, $M_{0\_12}$ for ESES-based current calibration purpose and the transistor widths of these sub-transistors were sized as an arithmetic sequence. Each sub-transistor is controlled by a pair of switch at the gate. For example, when ESES control signal $S_1 = 1$, the gate of $M_{0\_1}$ connects to the MSB biasing voltage and $M_{0\_1}$ contributes DC biasing current. When $S_1 = 0$, the gate of $M_{0\_1}$ is grounded. In this case, $M_{0\_1}$ is turned off and it only has negligible leakage current. A total of 6 sub-transistors out of the 12 sub-transistors are turned on, which makes 924 available design choices for each UCC during the ESES-based calibration process.

A cascode transistor $M_1$ is placed at the drains of the sub current sources for shielding the capacitance of that node. Similar to the SES-based CS-DAC design as shown in section 5.1.2 ([11]), amplitude error calibration is only performed on the UCCs of the 6-bit thermometer coded MSBs. The amplitude errors of the LSBs are optimized by upsizing LSBs' transistor channel length by 8 times. Therefore, MSBs and LSBs have different channel lengths and parasitic capacitance values which are not proportional to the current values. This cascode transistor $M_1$ helps to significantly lower down the impact of the non-linearity from the parasitic capacitance.

A switching pair $M_2$ and $M_3$ are placed on top of the cascode transistor $M_1$. They are switched by complementary digital inputs Din_p and Din_n which are coming from the preceding latch. Similar to the CS-DAC implementation presented in [48], one extra level of cascode transistors $M_4$ and $M_5$ are placed at the drains of $M_2$ and $M_3$. These two transistors $M_4$ and $M_5$ are thick-oxide transistors for interfacing the off-chip resistive loading with 1.8 V supply voltage and shielding the rest of the 1.0 V thin-oxide transistors ($M_{0\_X}$, $M_1$, $M_2$, $M_3$) from 1.8 V power supply. The extra cascode transistors further increase the output impedance of the UCC, thereby improving the linearity of the CS-DAC. They also shield the output nodes from the digital switching input Din_p and Din_p. Two bleeding current sources, $I_{bleed\_p}$ and $I_{bleed\_n}$, are added at the sources of $M_4$ and $M_5$ respectively to make $M_4$ and $M_5$ always ON, independent of the current of the UCC being steered to $M_4$ or $M_5$. Therefore, less output impedance change is observed for the two input states. As the output impedance becomes less input data dependent, better dynamic linearity can be achieved. This technique was first proposed in [48], and was becoming popular in recent years [47] [49].

For amplitude/timing error measurement, one extra current output path (through $M_6$ and $M_7$ to error sensing circuit) is created in parallel with normal current output path (through $M_4$ and $M_5$ to off-chip loading). This implementation is similar to the one designed in [46]. M4/M5 and M6/M7 are controlled by complementary control signals. It is noted that the control signals need to be 1.8 V logic for these thick-oxide cascode transistors.

### 5.3.2 Latch Design with ESES-based Tunable Delay and Duty Cycle

The differential latch topology adopted in this design mainly consists of clock buffers ($INV_1$, $INV_2$), stacked NMOS transistors ($M_1/M_3$, $M_2/M_4$) for clock and data inputs, cross-coupled inverters ($INV_3$, $INV_4$) for storage, and output buffers ($INV_5$, $INV_6$). As the data inputs only connect to NMOS transistors rather than a complementary NMOS/PMOS configuration, there is only pull down path at the inputs of the cross-coupled inverter pair $INV_{3,4}$. For example, when D = 1 and DN = 0, at the rising edge of the clock, node X is pulled down to ground through the path of $M_1/M_3$. But on the other side, $M_2/M_4$ does not provide complementary pull-up path. The rising edge of node Y purely depends on the pull-up path inside $INV_3$, which is triggered by the falling edge of node X. Therefore, the rising edge of Y has a systematic delay as compared to the falling edge of node X. This is the same for the delay between the rising edge of X and the falling edge of Y. As a result, waveforms of node X and node Y always have a low crossing point. And with one more inverter buffering stage, Q and QN can always have a high crossing point that is beneficial for NMOS based UCC design to prevent $M_1$ in Figure 5.10 from going into the triode region, which would slow down the current switching.

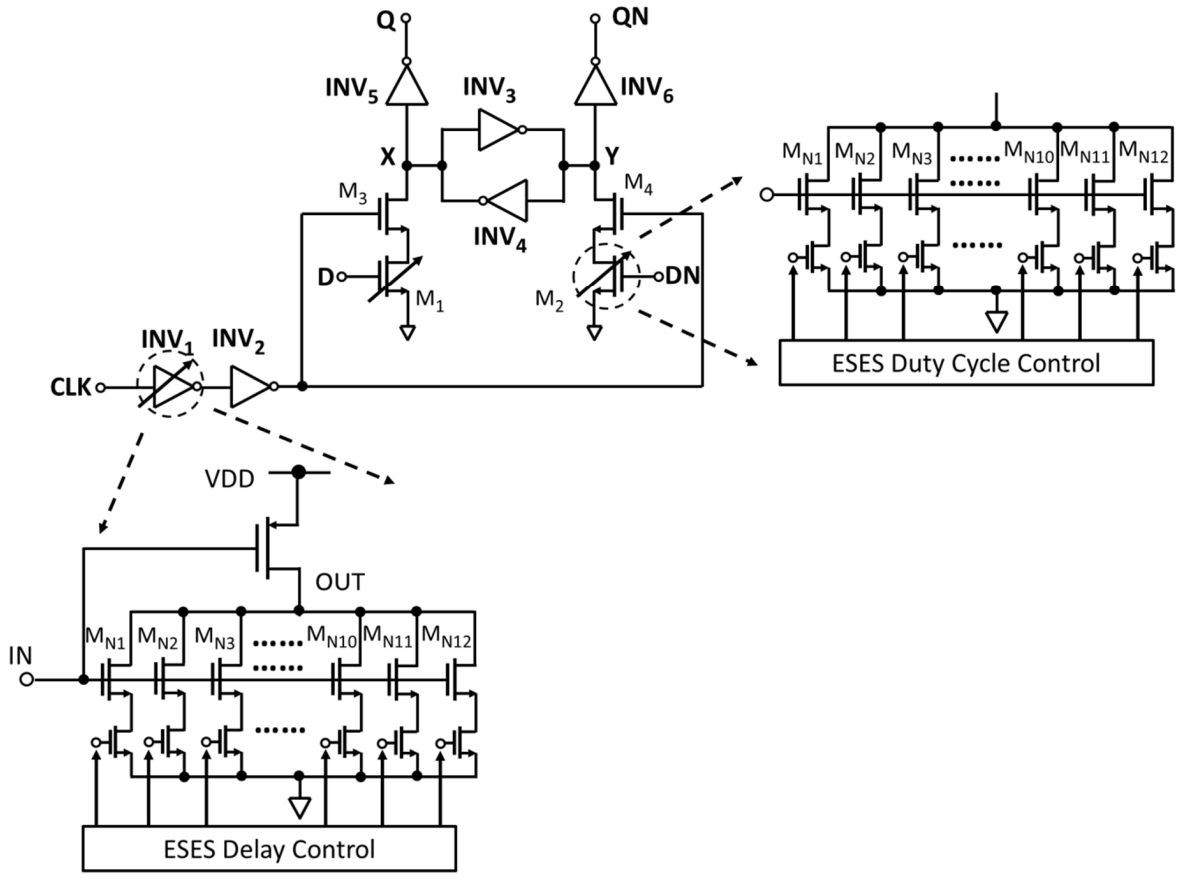

Figure 5.11 ESES-based latch design with tunable delay and duty cycle.

For delay calibration, the clock buffer $INV_1$ is employed with tunable delay by using ESES-based delay calibration method as presented in section 3.3. The delay tuning on $INV_1$ applies to both Q and QN outputs of the differential latch. For duty cycle calibration, we are calibrating $M_1$ and $M_2$ as shown in Figure 5.11. By tuning the driving strength of these two transistors, the clock to Q/QN delay for D = 0 and D = 1 cases can be tuned separately, thereby achieving calibration on Q/QN duty cycles. For both ESES-based delay error and duty cycle error calibration, the ESES design parameter N and K are chosen as 12 and 6 as shown in Figure 5.11, resulting in 924 design choices for each tunable component.

### 5.3.3 Error Sensing Circuit Design

The amplitude/timing error sensing circuit design is shown in Figure 5.12. The sensing circuit implementation is similar to the one designed in [46]. Two pairs of differential test currents are taken as inputs for the error sensing circuit. One differential test current is coming from one UCC, and the other differential test current is the summation of all 8-bit binary weighted LSBs plus one more extra LSB cell. The two pairs of the differential test currents have opposite polarities when connected to the error sensing circuit as shown in Figure 5.12. Therefore, the error sensing circuit is taking the difference of the two pairs of the differential test currents as inputs, which is the mismatch error between the two pairs.

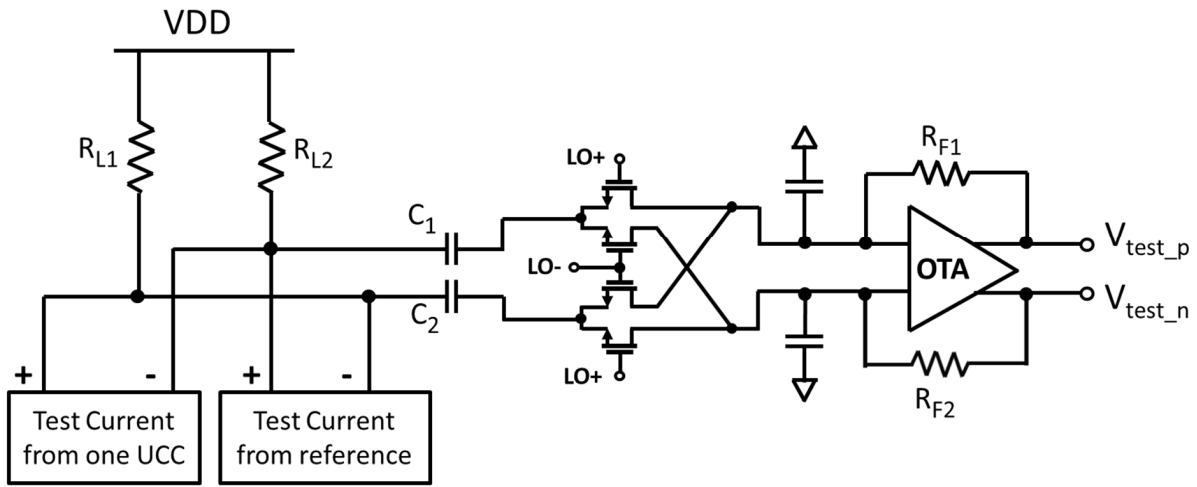

Figure 5.12 Amplitude/timing error sensing circuit.

Two resistive loads, $R_{L1}$ and $R_{L2}$, are connecting the test currents to the power supply. They are used to give DC current paths for the test currents. The AC component of the test currents, which is also the mismatch error between the two test currents, is mainly going through the AC coupling capacitor $C_1$ and $C_2$. A mixer stage is following the AC coupling capacitors. The mixer down-converts current mismatch signals into DC values. The modulation signals LO+ and LO- have different frequencies and phases settings for different calibration processes, namely amplitude, delay and duty cycle calibration processes, as described in section 5.2. A trans-impedance stage (TIA), which is realized as operational trans-conductance amplifier (OTA) with resistive feedback, is following the mixer stage and converting the current mismatch DC values into DC voltages. The OTA design is a typical fully-differential two-stage amplifier design with a Miller compensation capacitor. The circuit topology is similar to the one shown in Figure 4.8. The differential voltage outputs of the TIA are then analyzed by the off-chip error analyses unit. It is noted that even without any inputs from the test currents, this error sensing circuit can have non-zero DC outputs due to input mismatch of the OTA as well as flicker noise of the circuit. Therefore, before measuring the mismatch error between two test currents, an offset calibration process needs to be done to measure the DC outputs without any test current inputs. This value needs to be subtracted from the measured mismatch error of two test currents. This error sensing circuit is designed to have a 1 mV DC output for a 60 nA amplitude error, a 80 fs delay error and 160 fs duty cycle error at a measurement frequency $f_{meas}$ of 400 MHz.

## 5.4 Simulation Results

The ESES-based CS-DAC design is implemented in 28nm process and the core circuit occupies 0.12 mm$^2$. The layout of the CS-DAC core circuit is shown in Figure 5.13.

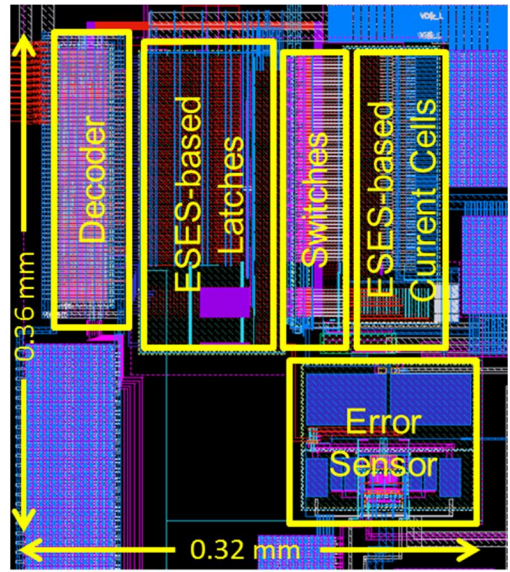

Figure 5.13 Layout of the CS-DAC core circuit.

The total output current of the CS-DAC is 20 mA. This consumes 36 mW at 1.8V power supply. The biasing circuit for the current cells consumes 2 mW at 1.0 V power supply. Each one of the 63 UCCs of the thermometer coded MSBs has a nominal current value of 312 μA. In Monte Carlo simulation, the standard deviation of the UCC current is 2.8 μA. Our ESES-based design method breaks the current source into 12 sub-transistors as shown in Figure 5.10 while selecting 6 out of them. The nominal current values of the 12 sub-current sources are shown in Table 5-2. The values are of an arithmetic sequence with an average of 52 μA and a common difference of 0.76 μA. The standard deviation of each sub-current source is approximately 1.1 μA, which is about 2% of the nominal sub-current source value. The calibration for each one of the 63 UCCs is performed by finding the optimal combination out of 924 possible combinations that has the closest value to a reference current that consists of all 8-bit LSBs and one more LSB.

Table 5-2 Nominal current values for 12 sub-current sources in UCC.

| Nominal Currents | 47.82 μA | 48.58 μA | 49.34 μA | 50.1 μA | 50.86 μA | 51.62 μA |
|---|---|---|---|---|---|---|
| | 52.38 μA | 53.14 μA | 53.9 μA | 54.66 μA | 55.42 μA | 56.18 μA |

System-level Monte Carlo simulation was performed for $10^4$ samples, with each sample representing one independent 14-bit CS-DAC design. Figure 5.14 shows the integral non-linearity (INL) plot for one typical sample before and after ESES-based amplitude error calibration. In this particular sample, the maximum INL ($INL_{max}$) is 10.5 LSB before calibration and 0.28 LSB after calibration. $INL_{max}$ for each of the $10^4$ samples is calculated before and after calibration. Figure 5.15 shows the $INL_{max}$ distributions for all samples before calibration. This distribution has a 99th percentile point as 20.5 LSB. Figure 5.16 shows the $INL_{max}$ distributions after calibration, and the 99th percentile point is reduced to 0.52 LSB. This means that if the yield target of the CS-DAC is 99%, then the $INL_{max}$ specification can be improved from 20.5 LSB to 0.52 LSB after ESES-based amplitude error calibration. Also from simulation, the maximum differential non-linearity ($DNL_{max}$) specification for 99% yield target can be reduced to 0.78 LSB from 10.8 LSB after calibration.

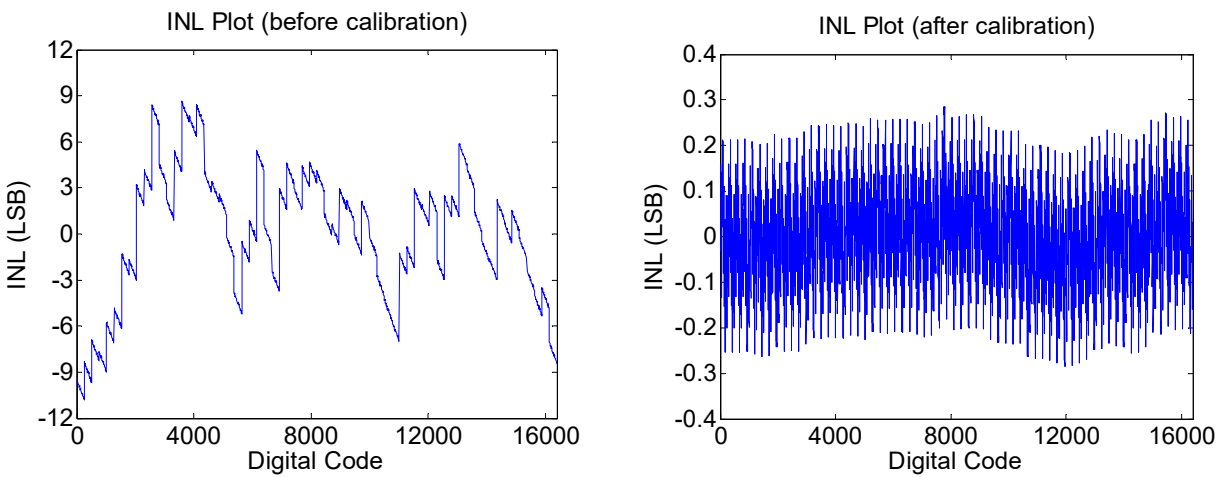

Figure 5.14 INL plot before and after ESES-based calibration for one typical sample.

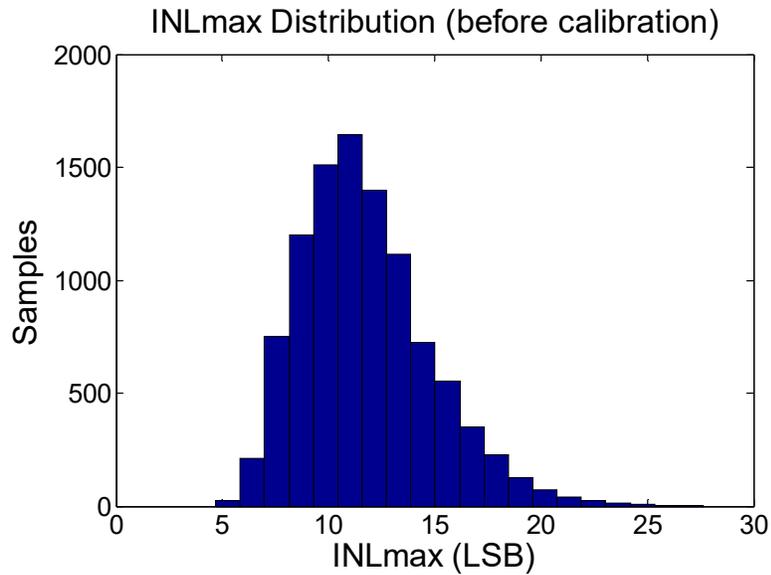

Figure 5.15 $INL_{max}$ distribution before ESES-based calibration.

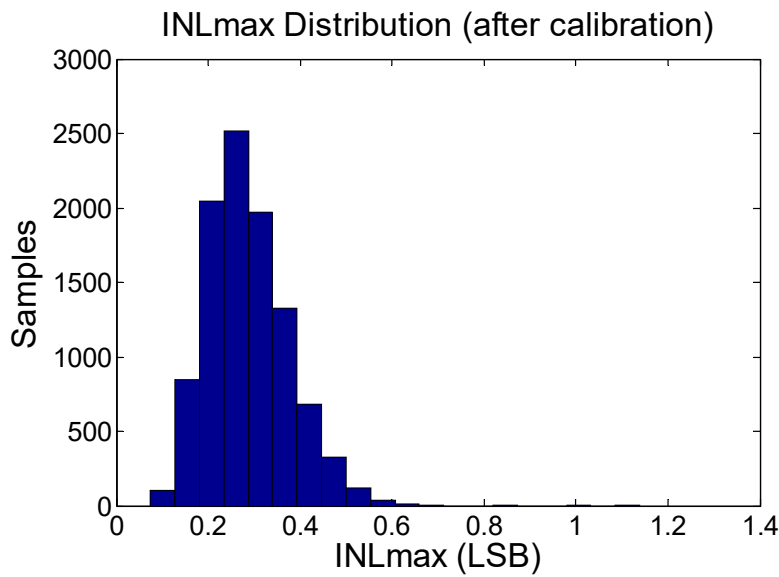

Figure 5.16 $INL_{max}$ distribution after ESES-based calibration.

For ESES and SES comparison purposes, we have also simulated calibration performance by using the SES method. After SES-based amplitude error calibration, for a 99% yield, the $INL_{max}$ specification is 5.6 LSB and the $DNL_{max}$ specification is 4.7 LSB. The $INL_{max}$ distributions for all $10^4$ samples are shown in Figure 5.17. As we can see from the histograms, the $INL_{max}$ distributions after SES-based calibration have a much longer tail as compared to ESES-based calibration. This is primarily due to the lower calibration success rate of the SES method, as discussed in section 3.1.3. The INL plot for one typical sample after SES-based calibration is shown in Figure 5.18. For a couple of UCCs among all 63 UCCs that are calibrated, SES cannot find a combination that has a current value close to the reference current. Therefore, the INL plot after SES-based calibration has a couple of steep steps, which degrades $INL_{max}$ performance.

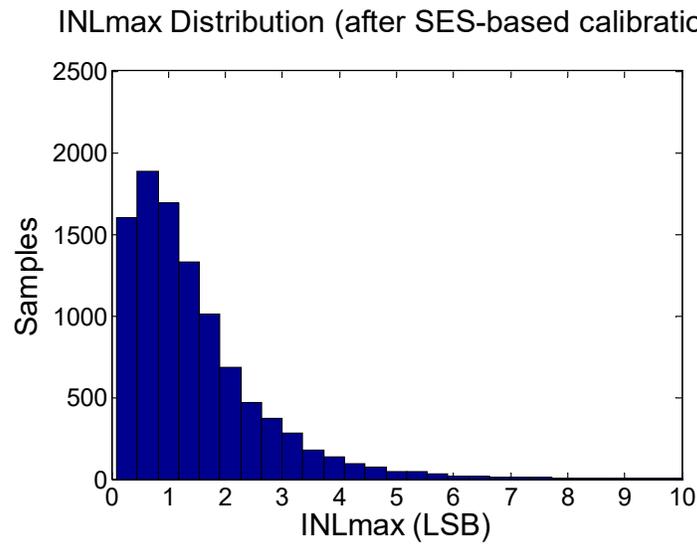

Figure 5.17 $INL_{max}$ distribution after SES-based calibration.

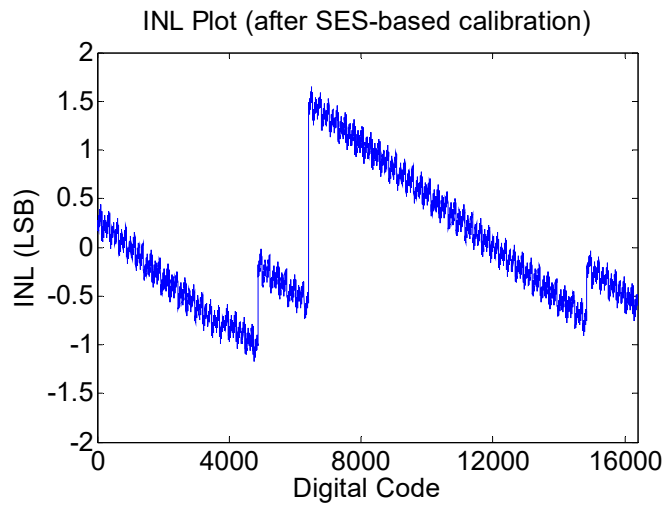

Figure 5.18 INL plot after SES-based calibration for one typical sample.

For the digital circuits in CS-DAC, the decoder is implemented with semi-custom digital circuit design flow by using Synopsys Design Compiler and Synopsys IC Compiler (ICC) tools. The maximum operating frequency reported by ICC is 3.2 GHz at 1 V power supply. The speed of the decoder is the ultimate speed limit for the data conversion rate for this CS-DAC design. The decoder also has a maximum power consumption of 22 mW at 1V power supply and 3.2 GHz clock frequency during post-layout simulation in Cadence Virtuoso tool.

The ESES-based latches were designed in Cadence Virtuoso like all other analog circuit components. Two stages of latches with a master-slave structure were employed for each UCC in the MSBs as well as for each unit in the binary weighted LSBs. The slave latch design is described in section 5.3.2 with ESES-based tunability for timing calibration. The master latch is a simplified version of the design as shown in Figure 5.11 without any ESES-based tunability employed, as the timing errors of the master latch can be corrected by the succeeding ESES-based slave latch. The entire latch array and clock buffers for the latches consumes a maximum power dissipation of 15 mW at 1V power supply and 3.2 GHz clock frequency during post-layout simulation in Cadence Virtuoso. Standard deviation of the delay errors for the 63 UCCs in 6-bit thermometer coded MSBs is simulated as 1.3 ps. The ESES-based delay calibration is performed on each UCC by exhausting all 924 combinations (we have N = 12 and K = 6) and choosing the one has the minimum delay error. Behavioral Monte Carlo simulation shows that after ESES-based delay calibration, the standard deviation of the delay errors of the UCCs can be reduced to 0.03 ps from 1.3 ps. For ESES-based duty cycle error calibration, it is not necessary to tune both $M_1$ and $M_2$ in Figure 5.11. For example, we can fix the selection of the sub-transistors in $M_2$, exhaust all 924 design choices for $M_1$, and find the one combination for $M_1$ that results in minimum duty cycle error. Behavioral Monte Carlo simulation shows that before calibration, standard deviation of the duty cycle errors is 1.8 ps, and after calibration, it reduces to 0.04 ps. In summary, the overall timing errors coming from the 6-bit MSB can be greatly reduced by our proposed ESES-based timing error calibration method. Therefore, according to the analyses in [46], we can expect a much better linearity performance of the CS-DAC at high frequency.

Table 5-3 Performance summary of the ESES-based CS-DAC with expected performance.

| | |
|---|---|
| Resolution | 14 bits |
| Max Sampling Rate | 3.2 GS/s |
| Full-scale Current | 20 mA |
| Output Swing | 1 V (differential) |
| Supply Voltage | 1.0 V/ 1.8 V |
| $INL_{max}$ (99% yield target) | 20.5 LSB (before)   0.52 LSB (after) |
| $DNL_{max}$ (99% yield target) | 10.8 LSB (before)   0.78 LSB (after) |
| Power Consumption | 38 mW (analog); < 37 mW (digital) |
| Core Area | 0.12 mm$^2$ in 28 nm CMOS technology |

Table 5-3 shows the performance summary of the ESES-based CS-DAC with expected performance from simulation. After calibration, we can achieve more than one order of magnitude linearity improvement for the CS-DAC.

## 5.5  Summary

An ESES-based CS-DAC design was presented in this chapter. Different from the previous SES-based implementation, this design calibrates both amplitude errors and timing errors of the CS-DAC, thereby optimizing linearity across the entire Nyquist band. With a wider calibration range of the ESES method, the current source spatial errors can be well covered by the calibration, resulting in a simple layout scheme that can benefit CS-DAC dynamic performance. Also as the ESES method has a higher calibration yield, less design choices are needed for redundancy and this can reduce the amount of calibration operations.

# 6  Conclusion and Future Work

As CMOS technology advances over the years, the minimum transistor size scales down exponentially. However, the transistor matching property has a slower pace of improvement. As a result, the analog circuit design cannot take the full benefit of the technology node scaling.

A design method called statistical element selection (SES) was proposed a couple of years ago for tackling the transistor matching issues. The idea is to break a transistor into a number of identically sized parts and then select a subset of them. This enables combinatorial redundancy, thereby achieving high calibration resolution in a small calibration range.

Although SES method was successfully applied to numerous circuit designs, including ADC and DAC designs, we find it short of being a more general calibration method due to its calibration range limitation. As the calibration range of SES method purely relies on random mismatch, the calibration range cannot exceed the random mismatch itself. This poses problems for the applications where multiple comparable variation sources exist in the design.

To overcome the shortcomings of the SES method, we proposed a new design method called extended statistical element selection (ESES). The ESES method intentionally skews the sizes of the sub-transistors, such that the calibration range becomes wider and deterministic by proper transistor sizing. Through case study, we also find that ESES not only provides wider calibration range as compared to SES, but also higher calibration yield when the calibration resolution is the same. Therefore, the new ESES method outperforms the SES method. With ESES, now we can perform calibration at a location where its random variation is not the dominant variation source in the design. This makes the ESES method become suitable for a broader range of applications for high resolution calibration purpose. We also proposed two types of ESES-based calibration applications. One is current source calibration, and the other is phase delay calibration.

We demonstrated utility of the ESES method in analog/RF designs through a design of harmonic rejection receiver. As harmonic rejection ratios are limited by gain errors and phase errors, we applied ESES-based calibration to both error sources to improve harmonic rejection ratios. The ESES-based design method was proven in this design to provide sufficiently large calibration range for covering various random variation sources in the receiver as well as provide high calibration resolution for achieving high matching accuracy. After calibration, we achieved best-in-class harmonic rejection ratios. However, the linearity performance of the receiver is inferior to other state-of-art receiver designs. This is mainly because of the receiver architecture that we chose for the implementation which has a large voltage gain in the RF band. For a future implementation, a mixer first or a low noise trans-conductance amplifier first wideband receiver architecture can be employed for the ESES-based harmonic rejection receiver design. We can also do ESES-based gain calibration at baseband, for example, calibration can be done through tuning baseband resistors. Therefore, potentially we can design a wideband receiver that is tolerant not only to harmonic interferences, but also other blockers with improved linearity performance.

We further presented a current steering D/A converter (CS-DAC) design that employs ESES-based calibration. As linearity of the CS-DAC suffers from amplitude errors with low frequency input and timing errors with high frequency input, we applied ESES-based calibration to both errors in order to achieve high linearity performance across the entire Nyquist band. Simulation results show that we can achieve more than one order of magnitude linearity improvement for the CS-DAC after calibration. As our design method is already proven by the receiver design, the simulation results give us confidence about the performance improvement. But future chip measurement results would be very helpful to confirm the benefit of performing ESES-based calibration in the CS-DAC.

Apart from the two circuit designs presented here, ESES-based calibration method can also be applied to more analog/RF circuits. For example, one can calibrate capacitor array in a SAR-ADC for achieving better linearity. Also for the comparator design inside flash ADC, with ESES method, one can now calibrate the loading of the comparator rather than the input differential pair of the comparator in order to reduce the parasitic capacitance at the inputs. More analog/RF applications can be further explored for taking the benefits of high matching properties after performing ESES-based high resolution calibration.